\pdfoutput=1
\documentclass[11pt,a4paper]{article}
\usepackage{setspace}
\usepackage{jheppub}

\RequirePackage[utf8]{inputenc}
\usepackage[utf8]{inputenc}

\usepackage{amsfonts}
\usepackage{comment}
\usepackage{tabu}
\usepackage{float}
\usepackage{amsthm}
\usepackage{slashed}
\usepackage[labelformat=simple]{subcaption}

\usepackage{listings}

\usepackage[font=small, labelfont=bf]{caption}
\usepackage{multirow}
\usepackage{float}
\usepackage{enumerate}
\usepackage{todonotes}

\newcommand{\nc}{\newcommand}
\nc{\beq}{\begin{equation}}
\nc{\eeq}{\end{equation}}
\nc{\bea}{\begin{eqnarray}}
\nc{\eea}{\end{eqnarray}}

\newcommand{\vol}{\mathcal{V}}

\newcommand{\K}{K\"{a}hler }
\newcommand{\Nsusy}{\mathcal{N}}
\newcommand{\antiD}{$\overline{D3}$ }
\newcommand{\antiDmath}{\overline{D3}}

\allowdisplaybreaks[2]
\numberwithin{equation}{section}

\usepackage[framemethod=TikZ]{mdframed}
\mdfsetup{skipabove=\topskip, skipbelow=\topskip}

\newcounter{equ}[section]

\newcounter{Boxequ}[section]

\title{On K3-fibred LARGE Volume Scenario with de Sitter vacua from anti-D3-branes}

\author[a]{Shehu AbdusSalam,}
\author[b]{Chiara Crinò,}
\author[c,d,e]{Pramod Shukla}

\affiliation[a]{\footnotesize Department of Physics, Shahid Beheshti University, Tehran, Iran}
\affiliation[b]{\footnotesize Dipartimento di Fisica, Università di Trieste, Strada Costiera 11, I-34151 Trieste, Italy and \\ INFN, Sezione di Trieste, via Valerio 2, I-34127 Trieste, Italy}
\affiliation[c]{\footnotesize ICTP, Strada Costiera 11, Trieste 34151, Italy}
\affiliation[d]{\footnotesize Department of Physics, University of Allahabad, Prayagraj 211002, India}
\affiliation[e]{\footnotesize Department of Physics, Unified Academic Campus, Bose Institute, \\ EN 80, Sector V, Bidhannagar, Kolkata 700 091, India}

\emailAdd{abdussalam@sbu.ac.ir}
\emailAdd{chiara.crino@phd.units.it}
\emailAdd{pshukla@jcbose.ac.in}

\abstract{
In the context of type IIB superstring compactifications on K3-fibred (weak) Swiss-cheese Calabi Yau (CY) orientifolds, we consider the realisation of de Sitter vacua obtained through the introduction of an \antiD- brane at the tip of a highly warped throat of Klebanov-Strassler type. Aiming to have a concrete global realisation, we perform a systematic search for the CY threefolds with $2 < h^{1,1} < 5$ arising from the Kreuzer-Skarke database, which satisfy the minimal requirements of being K3-fibred and suitable for moduli stabilisation within the LARGE Volume Scenario (LVS). In this context, after scanning the set of K3-fibred CY threefolds with a so-called diagonal del-Pezzo divisor needed for LVS, we realise that one of the main challenging requirements for having \antiD- brane uplifting is to find a suitable orientifold involution which can simultaneously result in a sufficient large $D3$ tadpole charge along with the presence of suitable $O3$-planes. In our detailed analysis (limited to) using the CY threefolds with small $h^{1,1}$, we observe that these topological requirements rule out most of the CY geometries leading to only few possibly suitable candidates for the purpose of \antiD- brane uplifting. Subsequently, we present a global model using one such explicit K3-fibred CY threefold with $h^{1,1}=4$ in which all the moduli can be consistently stabilised in a de Sitter minimum of the scalar potential.
}


\begin{document}

\maketitle

\bigskip



\section{Introduction}
\label{sec:intro}

The observed accelerating expansion of our universe is consistent with a description in terms of a de Sitter (dS) vacuum state. This is why a great amount of effort has been devoted over the years to the purpose of obtaining {dS space as a superstring compactification}; for a recent review see \cite{Cicoli:2018kdo} and references therein. Historically, we have witnessed that this task of finding stable dS solutions is very difficult, due to several theoretical obstructions\footnote{See, e.g., the famous Maldacena-Nu\~nez no-go theorem \cite{Maldacena:2000mw}.} which strongly constrain any attempt of explicit construction with the minimal set of ingredients at hand.

{One of the first mechanisms} for the realisation of dS vacua in type IIB orientifold compactifications was proposed in the so-called KKLT scenario \cite{Kachru:2003aw}\footnote{{We thank the referee for bringing our notice to \cite{Silverstein:2001xn,Maloney:2002rr} regarding earlier attempts on dS realizations.}}, based on the introduction of an \antiD-brane in the theory. Usually, the moduli arising from the low energy type IIB supergravity via performing the compactification on CY orientifolds are stabilized in an Anti-de Sitter (AdS) minimum with the inclusion of fluxes and instantons. Subsequently such AdS solutions can be `uplifted' to dS solutions by the inclusion of a variety of additional effects \cite{Cicoli:2018kdo}. In this regard, the main idea behind the KKLT proposal is that the presence of an \antiD-brane generates an additional term in the supergravity scalar potential, which, under certain conditions, can develop a (dS) minimum with positive energy.
Since \antiD- branes break supersymmetry, they are typically placed at the tip of a highly warped throat\footnote{We mention however that in \cite{Bento:2021nbb} a new metastable de Sitter solution within the 4d effective field theory was found also in a region of the parameter space corresponding to a `weakly-but-still-warped' throat that is a warped throat where, unlike the one we will consider here, the coupling between the conifold and the volume moduli is subdominant.}, so that the amount of supersymmetry breaking is small enough and the effective field theory is under control.
Since its original proposal, the KKLT mechanism has passed several consistency checks and survived many criticisms; for example, see \cite{Crino:2020qwk} for a brief review.

The presence of several non-trivial ingredients in the type IIB superstring compactifications, such as Calabi-Yau (CY) orientifolds,  three-form fluxes, branes, instantons etc., makes the construction of concrete dS models in this framework challenging, as many requirements need to be satisfied in order to combine all these ingredients in a consistent way. 
Moreover, there is a huge number\footnote{In \cite{Demirtas:2020dbm} it is proven that the upper bound for the number of topologically inequivalent Calabi-Yau hypersurfaces in toric varieties, arising from the Kreuzer-Skarke (KS) database \cite{Kreuzer:2000xy} is $N_{CY}^{max} \simeq 10^{428}$.}
of possible CY manifolds that can describe the compactified internal dimensions of the string theory models. 
Each of these can generically admit a large number of K\"ahler moduli, whose stabilisation is typically quite tricky. 
 So both theoretical and computational ``curse of dimensionality'' (related to the large number of variables and parameters in the scalar potential) difficulties can be foreseen \textit{en route} of the efforts for obtaining phenomenologically interesting models including dS vacua. For these reasons, within the literature, mainly examples with few K\"ahler moduli and selective base geometries were addressed, such as the already mentioned KKLT model, the Swiss-cheese LARGE Volume Scenario (LVS)~\cite{Balasubramanian:2005zx}, and many variants of these. However, some exceptions to the usual minimalistic approach of working with just a few number of moduli can also be found, e.g. see \cite{Denef:2005mm}, where $51$ \K moduli (and $3$ complex structure moduli) are explicitely stabilized, and \cite{Demirtas:2021nlu}, where vacua with $\mathcal{O}(100)$ \K moduli are analysed.
So far, there are many proposals of realising dS vacua from string theory model building attempts within the literature -- see for instance the constructions in the non-exhaustive references~\cite{ hep-th/0307160, hep-th/0309187, hep-th/0307084, Saltman:2004sn, Westphal:2006tn, AbdusSalam:2007pm, hep-th/0701154, Cicoli:2012fh, Retolaza:2015nvh, Gallego:2017dvd, Louis:2012nb, Rummel:2014raa, Braun:2015pza, Cicoli:2015ylx, Antoniadis:2019rkh, Cicoli:2017shd, Cicoli:2021dhg}. 

While constructing the phenomenology inspired models in a string theoretic framework, one usually makes several assumptions, particularly about the choice of the internal geometries. These assumptions translate into a series of requirements that a model has to satisfy in order to be consistent. In this context, there have been a great amount of efforts in global model building \cite{Cicoli:2011qg, Cicoli:2012vw, Cicoli:2013mpa, Cicoli:2013cha, Garcia-Etxebarria:2015lif, Cicoli:2016xae, Cicoli:2017shd, Cicoli:2017axo, AbdusSalam:2020ywo, Demirtas:2020ffz, Crino:2020qwk, Cicoli:2021dhg, Carta:2021kpk, Leontaris:2022rzj}, in order to make systematic  the simple model building proposals by finding CY threefolds satisfying such list of requirements, concerning e.g. the relevant divisor topologies. Furthermore, since the past decade there has been a surge in global model building studies because of the recent development of efficient techniques and packages such as \texttt{PALP} \cite{Kreuzer:2002uu, Braun:2011ik}, \texttt{SAGE} \cite{sagemath}, \texttt{cohomCalg} \cite{Blumenhagen:2010pv, Blumenhagen:2011xn}, and \texttt{CYTools} \cite{CYTools} which have helped in scanning the CY threefolds from the two major dataset collections, namely the Kreuzer-Skarke (KS) dataset \cite{Kreuzer:2000xy} and the so-called complete intersection CY (CICY) dataset \cite{Candelas:1987kf,Anderson:2017aux}. These packages have led to more pheno-friendly dataset collections depending on the properties of the CY threefolds such as those reported in \cite{Cicoli:2011it,Gao:2013pra,Altman:2014bfa,Cicoli:2018tcq,Carta:2020ohw,Altman:2021pyc,Gao:2021xbs,Carta:2022web,Crino:2022zjk,Shukla:2022dhz}. Equipped with these powerful expertise and the dataset, the analysis of simple concrete examples, has turned out to be crucial as it allows to address, in an explicit way, the many consistency conditions as well as to develop techniques that can subsequently be applied also to more involved and realistic string compactifications. 

Following this line of making explicit global constructions, a concrete model realising a dS vacuum within the minimal LVS framework with $h^{1,1}(CY)=2$ was constructed in~\cite{Crino:2020qwk}. 
The dS vacuum was realised by means of the introduction of a single anti-D3 brane on top of an O3-plane
at the tip of a highly warped throat of Klebanov-Strassler type~\cite{Klebanov:2000hb}.
It is important to notice that, though this is not the only admissible choice for the \antiD-uplifting prescription, it is particularly convenient. 
It is known, indeed, that the presence of the \antiD in an otherwise supersymmetric background breaks supersymmetry spontaneously.
The manifestly supersymmetric action describing this effect was derived e.g. in \cite{Kallosh:2014wsa, Bergshoeff:2015jxa, Kallosh:2015nia} (for flat and curved GKP backgrounds respectively), starting from a $\kappa$-symmetric brane in the supersymmetric background and applying the consistent supersymmetric orientifold conditions for the fields on the brane\footnote{{It should be recalled that an orientifold projection is required in order to preserve $\mathcal{N}=1$ supersymmetry in four-dimensions, starting from the type II $\mathcal{N}=2$ $10$d supergravity theory.}}. 
The remaining action is the so-called Volkov-Akulov action \cite{osti_4437296}, which has a non-linearly realized supersymmetry on a single $\mathcal{N}=1$ fermionic goldstino, with no bosonic superpartners. 
The same action is obtained by considering a single \antiD on top of an O3-plane both in the flat and curved background and, with respect to other possible orientifolded configurations (see \cite{Kallosh:2015nia}) has the advantage that all the scalars are projected out, and we only need to deal with the single fermionic goldstino.

In~\cite{Crino:2020qwk}, the analysis of the class of models corresponding to the configuration described above, led to the introduction of the various constraints that one needs to consider in order to obtain solutions in which all the approximations and (the known) perturbative expansions are under control.
Among these requirements, a special attention has been devoted to the ones related to the D3 tadpole cancellation condition, which turned out to be particularly constraining.
The stabilisation of the complex structure modulus parametrising the warped throat containing the \antiD, indeed, requires that background three-form fluxes be turned-on with typically large (positive) D3-charge contribution. 
This contribution must be compensated by a large negative D3-charge (coming, for the setups we consider, from orientifold planes and D7-branes).
An important ingredient to facilitate such large contribution turns out to be the presence of a particular D7-brane configuration, called Whitney brane \cite{Collinucci:2008pf, Collinucci:2008sq}\footnote{The importance of this configuration in relation to D3-tadpole cancellation conditions was recently studied in a systematic way in \cite{Crino:2022zjk}.}.

In the context of moduli stabilisation within the framework of LVS, another attractive construction corresponds to using orientifolds of the so-called K3-fibred CY threefolds \cite{Cicoli:2008gp}. These are different from the strong Swiss-cheese orientifold constructions used in realising the standard LVS, which involves only one so-called ``large" four-cycle in the CY volume expression. Having instead multiple ``large" non-local four-cycles, which is one of the salient features of the models based on the K3-fibred CY orientifolds, have various interesting applications towards phenomenology \cite{Cicoli:2011it,Cicoli:2016xae,Cicoli:2017axo,Cicoli:2008gp,Cicoli:2011yh,Cicoli:2011yy}. In particular, the advantage of this setup is that after having stabilized the ``small'' cycles and the overall volume (e.g. à la the minimal two-field LVS), there is still at least one massless (K\"ahler) modulus, which can be fixed via subleading effects and may have an important phenomenological role. As an example, the shallowness of this additional direction in the scalar potential would make the corresponding \K modulus a promising candidate as the inflaton field in the inflationary paradigm known as \textit{fibre inflation} \cite{Cicoli:2016xae, Cicoli:2017axo, Cicoli:2008gp}. Moreover, it has been observed for the CY threefolds arising from the reflexive polytopes of the KS database that the {fraction of K3-fibred CY threefolds increases with $h^{1,1}(CY)$, while the fraction of strong swiss-cheese CY threefolds decreases} \cite{Shukla:2022dhz}. For these reasons, as an extension to the proposal of \cite{Crino:2020qwk} which is based on strong Swiss-cheese CY orientifold with $h^{1,1} = 2$, in the current work we aim for the realisation of dS vacua (obtained by means of an \antiD- brane) in a K3-fibred CY orientifold within the LVS framework. We consider our analysis as a first step towards more concrete models including also inflation.

For this purpose, one needs a CY threefold with $h^{1,1}=3$ as the minimal requirement. Indeed a suitable manifold must contain at least a local rigid so-called diagonal del-Pezzo (dP) divisor, {to be defined in Section \ref{sec:pheno-insights}}, for having non-perturbative effects in the superpotential in order to realise LVS\footnote{A new class of LVS models have been proposed in \cite{AbdusSalam:2020ywo}, using instead a non-diagonal del-Pezzo divisors. This turns out to be possible thanks to some underlying symmetries in the CY threefold.}, and two ``large" four-cycle volume moduli. Moreover, as already mentioned, there is a considerably large list of technical/phenomenological requirements which one needs to take care of while building explicit global models: in particular, e.g., besides the physical consistency conditions listed in \cite{Crino:2020qwk}, there is a set of new constraints recently proposed in \cite{Junghans:2022exo, Gao:2022fdi, Junghans:2022kxg}. Furthermore,  the standard prescription of LVS usually fixes the overall volume of the CY along with those four-cycle volume moduli which appear in the non-perturbative superpotential. Hence, there are several K\"ahler moduli which generically remain unfixed in the standard framework. One needs therefore to invoke a different set of scalar potential terms sourced from other ingredients such as string-loop effects \cite{Berg:2005ja,Cicoli:2007xp,Berg:2007wt,Antoniadis:2018hqy,Antoniadis:2019rkh,Gao:2022uop} and higher derivative $F^4$-corrections appearing at $\mathcal{O}(\alpha'^3)$ \cite{Ciupke:2015msa}. For example, after considering the list of requirements into account for the CY threefolds with $2 < h^{1,1}(CY) < 5$, we find that the simplest suitable K3-fibred CY orientifold for the current purpose turns out to have $4$ \K moduli and only one diagonal rigid divisor for LVS. Therefore, in order to stabilise all of them we also include the relevant string loop corrections as well as the $F^4$-contributions.

Several difficulties concerning model building with multiple \K moduli, which are very common while using the orientifolds of the K3-fibred CY threefolds with several large four-cycles, have been addressed in recent years. Just to mention two important achievements, the new package \texttt{CYTools} \cite{CYTools} has provided new powerful methods for the construction of CY geometries with $h^{1,1}>10$; on the other side, in~\cite{AbdusSalam:2020ywo, Cicoli:2021tzt} a new approach to type IIB K\"ahler moduli stabilisation was proposed with a set of ``master formulae'' for the supergravity scalar potential for a generic number of moduli,  alleviating a fundamental difficulty for string theory compactifications scenarios with many moduli fields. These results set an important base for building a systematic framework for mining, out of the huge landscape of CYs and flux possibilities, the string theory vacua satisfying {physical constraints}. However, an immediate issue concerns the explicit stabilisation of moduli using the scalar potentials obtained from the above-mentioned master formulae, which generically depend on many parameters and contains several highly sensitive exponential terms. For the CY orientifold analysed in this paper, the minimisation of the scalar potential has been done, with some guidance from analytical results, using the simplicial homology global optimisation algorithm (\texttt{SHGO})~\cite{Endres2018} implemented in the scientific computing tool \texttt{SciPy}~\cite{2020SciPy-NMeth}. The current work can be considered as the first step towards the application of these algorithms to more generic scalar potentials for which we cannot have the same amount of analytic control as the one analysed here.

The paper is structured as follows: Section \ref{sec:review} is devoted to a brief review of the background material needed for this work. In particular, we summarise the recipe for realising dS string vacua via the introduction of an \antiD- brane.
In Section \ref{sec:subLeadCorr}, first we present the necessary ingredients of models based on K3-fibred CY orientifolds. Subsequently we describe our search for finding a suitable model, highlighting and motivating our requirements as well as the main challenges arising in the attempt to combine these requirements, at least for geometries with small $h^{1,1}$. In addition we briefly review the relevant sub-leading corrections to the scalar potential, which are needed to stabilise the moduli that are left massless by the LVS mechanism. In Section \ref{sec:explicit_model} we present, and explicitly analyse, a concrete K3-fibred CY orientifold with $h^{1,1}=4$, selected from the scan of the previous section. Next, in Section \ref{sec:modStab} we discuss the moduli stabilisation leading to dS minimum. Finally, in Section \ref{sec:conclusion} we outline our conclusions, as well as some possible future directions.


\section{A recipe for dS uplift using \texorpdfstring{\antiD}{anti-D3} in LVS}
\label{sec:review}

In this section we briefly review the construction of the supergravity scalar potential for the moduli and the uplift mechanism originally proposed in \cite{Kachru:2003aw}, for stabilising all the moduli in a dS minimum. 
We present the argument for a generic number $h^{1,1}_+$ of \K moduli, and we summarise the results obtained in the simplest LVS case, with $h^{1,1}_+=2$.
Moreover, we list the consistency conditions that one needs to impose in a concrete realisation in order to control all the (known) approximations. 

\paragraph{}

We restrict to type IIB orientifold compactifications with $h^{1,1}_-=0$ for which the moduli from the closed string spectrum reduce to: the axio-dilaton $S=e^{-\Phi}-iC_0$;\: $h^{1,1}=h^{1,1}_+$ \K moduli {$T_i=\tau_i + i \, \theta_i$, and $h^{1,2}_-$ complex structure moduli $U_\alpha$.}

\subsection{The scalar potential}\label{sec:scalarPot}

The $\Nsusy=1$ supergravity theory corresponding to the 4d low energy physics is described by the \K potential ${\cal K}$ of the following form:
\begin{equation}\label{eq:Kpot}
    \begin{split}
        \kappa_4^2 {\cal K} =& -\ln\left[-i\int \Omega\,(U_\alpha)\wedge\bar{\Omega}\,(\bar{U_\alpha})\right]\\
       & -\ln[S+\bar{S}]-2\ln\left[\vol\,(T_i, \bar{T_i})+\frac{\xi}{2}\left(\frac{S+\bar{S}}{2}\right)^{3/2}\right] \equiv {\cal K}_{cs} + {\cal K}_{Q}
    \end{split}
\end{equation}
where ${\cal K}_{cs}$ denotes the complex structure moduli dependent piece presented in the first line, and ${\cal K}_Q$ denotes the pieces arising from the quaternionic sector of the ${\cal N} =2$ theory; $\kappa_4=\sqrt{8\pi G_N}$ is the $4$-dimensional gravitational coupling. Further, the superpotential $W$ can be given as:
\begin{equation}\label{eq:superpotential}
	W=\int G_3\wedge\Omega\,(U_\alpha)+\sum_{i=1}^{n}A_ie^{-a_i T_i}\equiv W_{cs}+W_T\:.
\end{equation}
Here, we have included both the leading order $\alpha'$ corrections \cite{Becker:2002nn} to the \K potential and non-perturbative corrections to the superpotential \cite{Witten:1996bn} along with the usual Gukov-Vafa-Witten (GVW) flux superpotential \cite{Gukov:1999ya} denoted as $W_{cs}$. 
We consider exclusively the cases in which the latter is induced by either an Euclidean D3-brane $(a_i=2\pi)$ or a stack of D7-branes supporting a condensing $SO(8)$ gauge group $\left(a_i=\frac{\pi}{3}\right)$. 
We assume that $n\leq h^{1,1}_+$ divisors support a non-perturbative effect of this kind.

Furthermore, we recall that $\Omega$ is the nowhere vanishing holomorphic $(3,0)$-form depending only on the complex structure moduli; $G_3=F_3-i\, S\, H_3$ is the field strength for the RR and NSNS 2-form potentials $C_2,B_2\ (F_3=d C_2, H_3=d B_2)$ and $\vol$ is the overall volume of the CY threefold, which can be defined in terms of the \K form $J$ and hence of the triple intersection numbers $k_{ijk}$ and the volumes $t^i$ of the 2-cycles dual to the $\tau_i$:
\beq
	\vol=\frac{1}{3!} \int_{CY} J \wedge J \wedge J = \frac{1}{3!}\,k_{ijk}\,t^i(x)\,t^j(x)\,t^k(x), \qquad \tau_i = \frac{\partial {\cal V}}{\partial t^i}.
\eeq
Finally, $\xi=-\frac{\chi(CY)\,\zeta(3)}{2\,(2\pi)^3}$ where $\chi(CY)$ is the Euler characteristic of the CY threefold and $\zeta(3)\sim1.202$.

The \K potential and the superpotential determine the so-called $F$-term contribution to the 4D scalar potential given as
\beq\label{eq:vSUGRA}
	V =e^{\cal K}\,\left({\cal K}^{I \bar{J}}\left(D_{I} W\right)(D_{\bar{J}}\overline{W})-3\,|W|^{2}\right)\:,
\eeq
where ${\cal K}^{I\bar{J}}$ is the inverse \K metric and the \K covariant derivative is defined as $D_I W=\partial_IW+(\partial_I {\cal K})W$.

Turning-on the three-form $F_3/H_3$ fluxes in the CY background can generically allow to stabilise all the complex structure moduli and the axio-dilaton by the leading order effects.
These RR and NSNS harmonic three-form fluxes contribute positively to the total D3-charge with:
\beq\label{eq:genQflux}
	Q_{D3}^{\rm flux}=\frac{1}{(2\pi)^4\alpha'^2}\int F_3\wedge H_3\:.
\eeq
Moreover, due to the extension to p-forms of the Dirac quantisation condition \cite{Nepomechie1985, Teitelboim:1985yc} their integral over arbitrary closed 3-cycles $\Sigma_A$ must be integer:
\beq
	\frac{1}{(2\pi)^2\alpha'}\int_{\Sigma_A} F_3\in \mathbb{Z};\quad \frac{1}{(2\pi)^2\alpha'}\int_{\Sigma_A} H_3\in \mathbb{Z}\:.
\eeq
The part of the scalar potential \eqref{eq:vSUGRA} relative to the complex structure moduli and the dilaton is positive definite and {solutions to the $F$-flatness conditions $D_\alpha W=0$  (where the index $\alpha$ runs over the axio-dilaton and the complex-structure moduli only) leads to the no-scale vacua with non-zero K\"ahler moduli $F$-terms.} 
Fixing these moduli at their respective dynamically realised VEVs, fixes the corresponding term in the superpotential  as $\langle W_{cs}\rangle=W_0$ along with the dilaton at $e^{-\Phi_0}=\frac{1}{g_s}$ introducing a new parameter $g_s$ as the string coupling. 

The back-reaction of the fluxes on the geometry of the CY induces a deformation that can be described with the introduction of a warp factor $e^{2D(y)}$ (and of a corresponding conformal factor in front of the CY internal space):
\begin{equation}\label{eq:warpedMetric}
	ds^2=e^{2D(y)}g_{\mu\nu}(x)dx^\mu dx^\nu+e^{-2D(y)}\tilde{g}_{i\bar{\jmath}}(y)dy^i d\bar{y}^{\bar{\jmath}},
\end{equation}
where $g_{\mu\nu}$ is the metric of the 4d space-time, while $\tilde{g}_{i\bar{\jmath}}$ is the metric of the CY.
The regions of the internal metric where $e^{-2D}$ is large are called warped throats and turn out to play an important role in the uplifting mechanism under consideration.

While the flux complex structure moduli stabilisation leaves the \K moduli massless, the introduction of sub-leading corrections allows to stabilise them as well \cite{Kachru:2003aw, Balasubramanian:2005zx}. 
In the LARGE Volume Scenario \cite{Balasubramanian:2005zx}, which we consider here, the stabilisation of the \K moduli is due to the competition between the leading perturbative correction to the \K potential and the leading non-perturbative correction to the superpotential, respectively given as in Eq.~\eqref{eq:Kpot} and Eq.~\eqref{eq:superpotential}.

Assuming that the complex structure moduli and the axio-dilaton are supersymmetrically stabilised via $D_\alpha W=0$, we can compute the scalar potential \eqref{eq:vSUGRA} for the \K moduli as \cite{AbdusSalam:2020ywo}
\beq\label{eq:vLVSgen}
    e^{-{\cal K}}V_K=V_1+V_2+V_3\:,
\eeq
with
\begin{subequations}\label{eq:vLVS}
	\begin{equation}
		\begin{split}
			V_1&=\sum_{i,j=1}^{n} e^{-a_i\tau_i-a_j\tau_j} A_i \, A_j \,\cos(a_i\theta_i-a_j\theta_j)\left\{-4\left(\vol+\frac{\hat\xi}{2}\right)(k_{ijk}t^k)\,a_i\, a_j\right.\\
			&\left.+\frac{4\vol-\hat\xi}{(\vol-\hat\xi)}\,(a_i \tau_i)\,( a_j \tau_j) +\frac{4\vol^2+\vol\hat\xi+4\hat\xi^2}{(\vol-\hat\xi)(2\vol+\hat\xi)}( a_i \tau_i+a_j\tau_j)+\frac{3\,\hat\xi\,(\vol^2+7\vol\hat\xi+4\hat\xi^2)}{ (\vol-\hat\xi)(2\vol+\hat\xi)^2} \right\}\:,
		\end{split}
	\end{equation}

	\begin{equation}
	\begin{split}
		V_2&=\sum_{i=1}^{n}2\,e^{-a_i\tau_i} A_i \, |W_{0}|\cos(a_i\theta_i+\phi)\left\{\frac{4\vol^2+\vol\hat\xi+4\hat\xi^2}{(\vol-\hat\xi)(2\vol+\hat\xi)}\, a_i \tau_i+\frac{3\,\hat\xi\,(\vol^2+7\vol\hat\xi+\hat\xi^2)}{(\vol-\hat\xi) (2\vol+\hat\xi)^2}  \right\}\:,
	\end{split}
\end{equation}	
\begin{equation}
	\begin{split}
		V_3&= |W_0|^2\,\frac{3\,\hat\xi\,(\vol^2+7\vol\hat\xi+\hat\xi^2)}{2\, (\vol-\hat\xi) (2\vol+\hat\xi)^2}\:,
\end{split}
\end{equation}	
\end{subequations}
where $\hat\xi=\frac{\xi}{g_s^{3/2}}$, we have replaced $W_0 = |W_0|\, e^{i\phi}$, and for simplicity we have taken the $A_i$'s to be positive reals.
This master formula is quite generic and depends simply on the ansatz for the K\"ahler potential and the superpotential in Eq.~\eqref{eq:Kpot} and Eq.~\eqref{eq:superpotential}. A generic detailed analysis of the derivatives of this scalar potential in searching vacua is hard to perform analytically, and it can lead to AdS solutions \cite{Kachru:2003aw,Balasubramanian:2005zx} as well as metastable dS solutions \cite{Balasubramanian:2004uy, AbdusSalam:2007pm}.
It is important to note that in the large volume limit, while the scalar potential for the complex structure moduli scales as $\frac{1}{\vol^2}$, the K\"ahler moduli dependent piece $V_K$ scales as $\frac{1}{\vol^3}$ due to the so-called no-scale structure.
The two-step procedure of moduli stabilisation described above is therefore justified in the sense that one can stabilise various (types of) moduli by simply considering the scalar potential terms order by order in an iterative manner via using the inverse of the overall volume ${\cal V}$ as the expansion parameter.

\subsection{dS vacua from \texorpdfstring{\antiD}{anti-D3} branes}
\label{sec:dSuplift}

One of the most studied mechanisms to obtain a theory in which all moduli are stabilised in a dS minimum prescribes the introduction of an \antiD brane at the tip $(y=0)$ of a warped throat \cite{Kachru:2003aw}.
Its contribution to the scalar potential is \cite{KKLMMT}
\begin{equation}\label{eq:upliftTerm}
    V_{\antiDmath}=2\,T_3\,e^{4D(0)}\simeq\frac{e^{4A_0}}{\vol^{4/3}}\,m_p^4\:,
\end{equation}
where $T_3$ is the tension of the brane and $m_p$ the (reduced) Planck mass. 
Moreover, we have defined $e^{-4D}=1+\frac{e^{-4A}}{\vol^{2/3}}$, from which it is clear that a strongly warped throat is characterised by $e^{-2A}\gg\vol^{1/3}$.
The term \eqref{eq:upliftTerm} is positive definite and its magnitude is governed by the warp factor, which in turn is fixed by background fluxes. 
Hence, it can serve as an uplifting term, provided that we tune the fluxes appropriately.
Warped throats usually arise in correspondence with conifold singularities deformed by the blowing-up of an $S^3$: the warp factor is therefore controlled by the complex structure modulus $Z$ parametrising such deformation, which is stabilised by turning on $M$ units of $F_3$ flux over the $S^3$ and $K$ units of $H_3$ flux over the dual cycle \cite{Giddings:2001yu}.

When the \antiD is placed on top of an $O3$-plane at the tip of the warped throat, the degrees of freedom of the system are fully captured by the introduction in the supergravity theory of a nilpotent goldstino (that is a superfield $X$ such that $X^2=0$, whose only propagating field is the goldstino) \cite{Kallosh:2015nia}.
We consider, in particular, the case in which the warped throat is equipped with two $O3$-planes sitting at the opposite poles of the $S^3$ (whose dimensions are fixed when the complex structure modulus $Z$ is stabilised), so that one can place a D3-brane on top of one of the O3-planes and an \antiD on top of the other.
This configuration ensures that the D3/	\antiD system does not contribute to the total D3-charge (the two branes have opposite D3-charge) and that there is no perturbative decay channel between the two branes (they are stuck at the $O3$ loci) \cite{Garcia-Etxebarria:2015lif}.

As it was noticed in \cite{Bena:2018fqc,Blumenhagen:2019qcg} in a strongly warped regime, the uplift term \eqref{eq:upliftTerm} scales in the same way as the term in the scalar potential which is responsible for the stabilisation of the complex structure modulus $Z$.
Hence, stabilising $Z$ before the introduction of the uplifting term in \eqref{eq:upliftTerm} (i.e.~together with the other complex structure moduli) might not be generically consistent.
For this reason, we consider this modulus separately with respect to the other complex structure moduli and we stabilise it explicitly. This subsequently means that we redefine the parameter $W_0$ so that now it indicates the VEV of the flux superpotential for all the complex structure moduli and axio-dilaton, except the $Z$ modulus.

The supergravity theory for the \K moduli $T_i$, the complex structure modulus of the throat $Z$ and the nilpotent goldstino $X$ is described by the following \K potential and superpotential, valid at the leading order near the minimum $D_ZW=0$:
\beq
	{\cal K}={\cal K}_{LVS}+\frac{c'\xi'|Z|^{2/3}}{\vol^{2/3}}+\frac{X\bar{X}}{\vol^{2/3}}
\eeq 
\beq
	W=W_0 +\sum_i A_i\,e^{-a_i T_i} -\frac{M}{2\pi i}\,Z\,(\log Z-1)+i\frac{K Z}{g_s}+\eta \,X
\eeq
where $c'\approx 1.18$ is a constant whose value was computed in \cite{Douglas:2007tu}; $\eta=\frac{Z^{2/3}i S}{M}\frac{c''}{\pi}$ \cite{Dudas:2019pls}, with $c'' \,$ 
$\approx 1.75$ \cite{Conlon:2005ki}.
The integer fluxes $\{M,K\}$ on the internal cycles of the conifold contribute to the total D3-charge with (see Eq. \eqref{eq:genQflux}):
\beq
	Q_{D3}^{\rm flux}=MK\:.
\eeq
A large hierarchy of masses \cite{Giddings:2001yu} as well as the need of a small uplift constrain the conifold modulus to take a small VEV given as:
\beq
	|Z|\ll1\:.
\eeq
Under this assumption it can be shown that the new scalar potential reduces to \eqref{eq:vLVSgen} plus an additional term:
\beq\label{eq:vLVSup}
	V\approx V_{\rm LVS}+\frac{\zeta^{4/3}}{2c' M^2\vol^{4/3}}\left[\frac{c'c''s}{\pi}+\frac{M^2\sigma^2}{4\pi^2}+\left(s K+\frac{M}{2\pi}\log{\zeta}\right)^2\right]
\eeq
where we used $Z=\zeta e^{i\sigma}$. This way, the complex structure modulus of the conifold is stabilised at:
\beq\label{eq:zMin}
    \sigma_0=0; \quad \zeta_0=e^{-\frac{2\pi K}{g_s M}-\frac{3}{4}+\sqrt{\frac{9}{16}-\frac{4\pi c' c''}{g_s M^2}}}\:.
\eeq
As observed in \cite{Bena:2018fqc}, the presence of a square root in \eqref{eq:zMin} implies that we need to impose an additional constraint on the solution, in order to avoid the destabilisation of the scalar potential as a consequence of the uplift mechanism: $g_s M^2\geq \frac{64\pi}{9}c'c''\approx 46.1\:.$
However, this constraint has been recently significantly relaxed in \cite{Lust:2022xoq}. 
Here, indeed, it is proven that the scalar potential for the complex structure modulus $Z$ far from the minimum  (in particular in the IR limit $Z\rightarrow 0$) is not well described by \eqref{eq:vLVSup}.
A more accurate computation of such scalar potential results in a milder constraint, where a possible destabilisation due to the presence of the \antiD is expected only if $g_s M^2\approx 1$. 
Nonetheless, this is not expected to significantly change our results, as there is also, as we will review in Section~\ref{sec:list-of-requirements}, the more constraining requirement $g_s M\gg1$, which still needs to be satisfied.
However, the scalar potential \eqref{eq:vLVSup} and the corresponding expressions for the moduli at the minimum \eqref{eq:zMin} can be considered to be a good approximation of the scalar potential near the minimum, which is the region we are interested in exploring, and we will utilise them for our purpose.

\subsection{An explicit example with \texorpdfstring{$h^{1,1}=2$}{h11=2}}

In order to make the above arguments more concrete, this section is devoted to a brief review of the results obtained in the simplest possible LVS configuration, in which there are only two \K moduli and the volume of the CY can be written in the strong Swiss-cheese form:
\bea
\label{eq:swiss-cheese-volume}
\vol=\tau_b^{3/2}-\kappa_s\tau_s^{3/2}\:,
\eea
where $\kappa_s$ is a constant depending on the intersection numbers of the CY threefold, and the K\"ahler cone conditions ensure that ${\cal V}$ is positive definite.
As we will see, the formulae of this section will be also useful for the analysis of the explicit model with $h^{1,1}=4$, addressed in Section \ref{sec:explicit_model}.

\paragraph{}

The usual LVS scalar potential after being appended with the $Z$-modulus dependent pieces reads as follows:
\begin{eqnarray}\label{eq:v2moduli}
           	V &=&\frac{4 a^2\, A^2 \,g_s \sqrt{\tau_s} e^{-2 a \tau_s}}{3 \kappa_s\mathcal{V}}+\frac{2 a\, A \, g_s \tau_s \,|W_0| \, e^{-a \tau_s}}{\mathcal{V}^2}\cos(a\theta_s+\phi)+\frac{3 |W_0|^2 \xi}{8 \sqrt{g_s}\vol^3} +  \nonumber \\  \\
	&& +\frac{3 \zeta ^{4/3}}{8 \pi^2 c' \vol^{4/3}} \left[\frac{8\pi c' c''}{3g_s M^2}+\frac{8\pi^2 K^2}{3 g_s^2 M^2}+\frac{8\pi K}{3 g_s M}\log\zeta+\frac{2}{3}\log^2\zeta\right] \:. \nonumber  
\end{eqnarray}
In order to obtain a dS minimum, the term proportional to $|W_0|^2$ has to compete with the one $\propto\frac{\zeta^{4/3}}{\vol^{4/3}}$, which means that we need to tune the fluxes so that
\beq
    \frac{\zeta^{4/3}}{\vol^{4/3}}\sim\frac{|W_0|^2}{\vol^3} \Leftrightarrow \zeta^{4/3}\sim \frac{|W_0|^2}{\vol^{5/3}}\:,
\eeq
which is in agreement with the requirement of having $\zeta$ small.
The moduli $\sigma$ and $\theta_s$ are decoupled from the other moduli and they can be stabilised at $\sigma=0$ and $\theta_s=\frac{\pi-\phi}{a}$, therefore we can substitute their stabilised values directly in \eqref{eq:v2moduli}.

By computing the derivatives of the scalar potential, one finds the values of the moduli $\tau_b, \tau_s$ at the minimum, which, introduced in  \eqref{eq:v2moduli} allow to evaluate the scalar potential at the minimum:
\begin{equation}\label{eq:vMin}
	V_{min} = \frac{5 q_0 \zeta^{4/3}}{9 \tau_b^2}-\frac{6 \, |W_0|^2\, g_s \kappa_s\tau_s^{3/2}}{\tau_b^{9/2}}\frac{(a \tau_s-1)}{(4a\tau_s-1)^2} \:,
\end{equation}
where
\beq\label{eq:q0}
q_0=\frac{3}{16\pi^2 c'}\left(\frac{3}{4}-\sqrt{\frac{9}{16}-\frac{4\pi c' c''}{g_s M^2}}\right)\:.
\eeq

\subsection{Consistency conditions for the dS minimum}
\label{sec:list-of-requirements}

In this section we review the conditions that one needs to impose on the moduli VEVs at the minimum,  as well as on the parameters of the model $\{W_0, g_s, a_i, A_i, M,K,\chi(X)\}$, in order to have a consistent solution. 
We mention here only the most relevant ones, which are useful for our analysis, referring to \cite{Crino:2020qwk} for more details. 

First, the moduli need to be stabilised in a region in which the overall volume of the CY is large enough (so that we have control over the $\alpha'$ expansion) and the string coupling is small (so that we can trust the perturbative string approximation):
\beq
	\vol\gg\frac{\xi}{g_s^{3/2}}\gg1 \:; \qquad g_s\ll1\:.
\eeq			
Moreover, the consistency of the supergravity approximation requires the single $4$-cycles to be stabilised within the \K cone\footnote{This condition was automatically satisfied in the case analysed in \cite{Crino:2020qwk}, with only $2$ \K moduli in LVS. However, we need to impose it explicitly when dealing with more generic configurations.} and with large enough values: $\tau_i\gg1$ \footnote{We recall that in order to avoid multi-instanton effects, one should also impose $a_i\tau_i\gg1$ for the moduli contributing to the non-perturbative superpotential \eqref{eq:superpotential}. 
However, here we consider only cases in which $a_i\gtrsim1$, hence this condition is automatically satisfied for large values of $\tau_i$.}. The volume of the CY is also constrained by the requirements related to the presence of an \antiD- brane at the tip of a warped throat: 
\beq
\label{eq:two-rhos}
	\vol^{2/3}\rho\ll1\;; \qquad \rho^{1/4}\vol^{2/3}\gg1
\eeq
where we have defined the warp factor $\rho\equiv q_0 \zeta^{4/3}$.
The first condition of (\ref{eq:two-rhos}) corresponds to the need for a highly warped throat, while the second one ensures that the massive string states of the \antiD are still negligible with respect to $m_{3/2}$, despite the fact that they are redshifted to lower masses \cite{Cicoli:2012fh}.

The 4d description provided by the low-energy supergravity can be trusted if the several energy scales are such that the hierarchy $m_{3/2},\ m_{moduli}\ll M_{KK}\lesssim M_s\ll M_p$ holds. 
These requirements are automatically satisfied in the large volume limit, with the exception of $M_{KK}\gg m_{3/2}$, which, when expressed in terms of the parameters of the model, imposes an upper bound on $W_0$ \cite{AbdusSalam:2020ywo, Cicoli:2013swa}: 
\beq\label{eq:Wconstr}
	\sqrt{\frac{\kappa}{\pi}}|W_{0}|\ll\vol^{1/3}
\eeq
with $\kappa=\frac{g_s e^{K_{cs}}}{2}$. Actually this constraint comes from the requirement of having control over the derivative expansion of the effective field theory \cite{Cicoli:2013swa}. The expansion parameter in this case (a supersymmetric field theory with broken supersymmetry) depends on the coupling $g\sim M_{KK}/M_P\sim 1/\vol^{2/3}$ of the heavy states (the KK modes in our case) to the light states and on the normalised magnitude of the F-term $F\simeq M_P^2 W_0/\vol$. It is estimated to be \cite{Cicoli:2013swa}: $\epsilon=\frac{g F}{M_{KK}^2}\sim \frac{W_0}{\vol^{1/3}}$, from which follows the constraint \eqref{eq:Wconstr}. Finally, the need to trust the KS solution imposes a lower bound on the flux number $M$, which basically arises from the demand of a large enough size of the $S^3$ at the tip of the throat, and can be given by the following constraint:
\beq\label{eq:fluxCons}
	g_sM\gg1\:.
\eeq
As we mentioned before, indeed, the constraint $g_s M^2\gtrsim 46.1$ \cite{Bena:2018fqc} has been recently proven not to be needed anymore, as the exact scalar potential for the complex structure modulus $Z$, is actually more stable than what was expected up to now \cite{Lust:2022xoq}.
On the other side the possible values of the fluxes $M,K$ are constrained by the D3-tadpole cancellation condition which, in a setup in which we explicitly stabilise only one complex structure modulus, reads:
\beq
	|Q_{D3}^{O3/O7/D7}|>MK\:,
\eeq
where $Q_{D3}^{O3/O7/D7}$ is the negative D3-charge contribution coming from $O3$-planes, $O7$-planes and D7-branes.  

\paragraph{}

In addition to the ones reported before, a series of new requirements has been recently proposed in \cite{Junghans:2022exo, Gao:2022fdi, Junghans:2022kxg}.
These constraints derive from the need to ensure that all possible corrections that can be added to the minimal scalar potential \eqref{eq:v2moduli} are actually negligible.
The main point of these papers is, indeed, that even corrections that are usually suppressed in the \textit{off-shell} scalar potential, may turn out to be relevant when computing \textit{on-shell} quantities such as the value of the scalar potential at the minimum as well as the stabilised moduli.
In particular, some types of corrections, when present, appear not to be suppressed by any small parameter, so that they can be neglected only if certain combinations of the parameters defining the geometric setup are small enough. 
This is the case, e.g. of the correction coming from the one-loop field redefinition of the `small' \K modulus, which appears in the form $\tau_s\rightarrow\tau_s+C_i^{\rm log}\ln \vol$ \cite{Conlon:2010ji}, which can be neglected only if it is possible to choose a geometric setup such that the condition
\beq\label{eq:CsLogCorr}
	\frac{\xi^{2/3}a_s^2|C_s^{\rm log}|}{(2\kappa_s)^{2/3}}\ll g_s
\eeq
is satisfied. 

Other corrections, instead, can be in principle suppressed by suitable choices of the parameters, even though the author of \cite{Junghans:2022exo} claims that it is impossible to make all these corrections negligible at the same time. 
The constraints to be imposed in order to ensure that these corrections are suppressed are schematically reported in the form of $5$ quantities $\lambda_i\ (i=1,...,5)$ which are required to be small \cite{Junghans:2022exo}:
\beq
	\lambda_i\ll 1\:.
\eeq
However, it is important to note that \textit{all} these constraints have an implicit dependence on the (stabilised) complex structure and open string moduli VEVs which can be only quantified in analysis where dynamics of all the moduli are considered in an explicit manner. This is indeed a highly non-trivial task, especially due to the presence of a multi-field scalar potential having several parameters involved.
In the analysis of \cite{Junghans:2022exo}, the underlying assumption appears to be the fact that all the dependences parametrised by certain ``effective" constants $C_\circ^\circ$ (see e.g. $C_s^{\rm log}$ in \eqref{eq:CsLogCorr}), are generically taken to be $\mathcal{O}(1)$. However, in explicit constructions, a priori nothing forbids them to be smaller by a few orders. For example, one such incidence can be considered from the proposal of exponentially low VEV for the flux superpotential parameter $W_0$ \cite{Demirtas:2019sip} which may be taken to be against the usual notion of what is a tuned value and what is natural for $W_0$. 
Moreover, let us illustrate this point by considering another example, and taking one of the corrections, namely the so-called `log-redefinitions' of the chiral coordinates ($T_i$). Such corrections have been studied in the literature several times, and also in the context of LVS, e.g. in \cite{Conlon:2010ji} where it was argued to be present in a very specific case, for example only for a rigid local divisor and not for the so-called big divisor of the LVS. If such a shift is considered in the small divisor volume modulus $\tau_s$, one gets the following \cite{Conlon:2010ji},
\bea
& &  \tau_s^{\rm new} = \tau_s^{\rm old} - {\cal C}_s^{log} \, \ln{\cal V}.
\eea
For such a correction which is sourced at 1-loop level, it has been argued that the coefficient ${\cal C}_s^{log}$ is to be small, possibly as small as $10^{-3}$ \cite{Conlon:2010ji} in order to {prevent the logarithmic shift from competing with the classical term.} Although the explicit computations of such coefficients are yet to be known, it could be fair to argue that there is no reason to enforce the coefficient $C_s^{log}$ to be ${\cal O}(1)$. 
In the following we will compute the constraints introduced in \cite{Junghans:2022exo, Gao:2022fdi, Junghans:2022kxg} (up to the unknown factors $C_\circ^\circ$), and we will assume that many of these can be tuned small in an appropriate flux vacuum. Investigating further the existence of some natural tuning mechanism suppressing the entire series of even higher order corrections would be interesting, but it is beyond the scope of this paper.


\section{K3-fibred model and sub-leading corrections}
\label{sec:subLeadCorr}

\subsection{Phenomenological insights of K3-fibrations}
\label{sec:pheno-insights}

One of the sub-classes of LVS moduli stabilisation with several phenomenological applications corresponds to a model of type IIB compactification over a manifold whose volume is controlled by more than one \K moduli, and in particular, it can be written in the so-called \textit{weak Swiss cheese} form:
\beq
	\vol=f_{3/2}(\tau_j)-\sum_{i=1}^{N_{\rm small}}\kappa_i\tau_i^{3/2}; \quad \text{with } j=1,...,N_{\rm large}\:.
\eeq
Here $N_{\rm small}$ \K moduli are associated to rigid cycles supporting non-perturbative effects, while the remaining $N_{\rm large}=h^{1,1}_+-N_{\rm small}$ \K moduli are combined in the homogeneous function of degree $\frac{3}{2}$, $f_{3/2}$, so that, in the large volume limit $\vol\simeq f_{3/2}$. For the case of $h^{1,1} = 2$ one has the trivial case corresponding to the minimal LVS construction as $f_{3/2} = \tau_b^{3/2}$ leading to the CY volume ${\cal V}$, given by Eq.~(\ref{eq:swiss-cheese-volume}). This shows that the next simple construction of this type will correspond to a CY threefold with $h^{1,1} = 3$ having the following volume form:
\beq
\label{eq:K3-volumeform}
	\vol= \tau_b \sqrt{\tau_f} - \kappa_s \, \tau_s^{3/2}\:.
\eeq
This structure in the volume form may be important from the phenomenological point of view. In particular after having stabilised the small cycles and the overall volume in LVS, there is always at least one massless modulus left, which can be stabilised by sub-leading effects such as the ones presented in the next section. Just to make an example, the shallowness of this additional direction in the scalar potential makes the corresponding \K modulus a promising candidate as, e.g. the inflaton field in the inflationary paradigm known as Fibre Inflation \cite{Cicoli:2008gp, Cicoli:2016xae,Cicoli:2017axo}.

Interestingly it turns out that a volume form such as (\ref{eq:K3-volumeform}) can be realised using a K3-fibred CY threefold. The simplest reason for this to happen lies in the fact that for a divisor $D\equiv$ K3 of a CY threefold the following relations hold:
\bea
& & \int_{X_3} \hat{D} \wedge \hat{D} \wedge \hat{D} = 0 \quad \Longleftrightarrow \quad  h^{1,1}(D) = 10\, h^{0, 0}(D) - 8\, h^{1,0}(D) +10\, h^{2,0}(D)\,,
\eea
where $\hat{D}$ denotes the $(1,1)$ homology class dual to the K3 divisor $D$.
This shows that for a K3 divisor there are no non-vanishing self-cubics in the intersection polynomial.
Subsequently, it directly follows from a theorem of \cite{OGUISO:1993, Schulz:2004tt} that if the CY intersection polynomial is linear in the homology class $\hat{D}$ corresponding to a divisor $D$, then the CY threefold has the structure of a K3 or a ${\mathbb T}^4$ fibration over a ${\mathbb P}^1$ base. For this to happen one needs to check that not only self-cubics but also self-quadratics vanish for the K3 divisor $D$ of the CY threefold, i.e.
\bea
\int_{X_3} \hat{D} \wedge \hat{D} \wedge \hat{\cal{D}} = 0, \quad \qquad \quad \forall \, \hat{\cal D}.
\eea
An immediate implication of this requirement for a CY with $h^{1,1} = 2$ is that if it is K3-fibred, the intersection polynomial can always be reduced to the following form for a suitable choice of divisor basis:
\bea
I_3 = k_{122} \, D_1\, D_2^2, 
\eea
where the divisor $D_1=$ K3 (and therefore its self-cubics and self-quadratic pieces are trivial), while $k_{122}$ denotes the corresponding triple intersection number. Now it is easy to convince that the above intersection polynomial leads to the volume of a K3-fibred CY threefold of the form ${\cal V} \propto \tau_2 \sqrt{\tau_1}$. Subsequently, after promoting the model to the case of $h^{1,1} = 3$, one can introduce an additional so-called diagonal del-Pezzo divisor as needed for LVS. {Here we note that a del-Pezzo surface $dP_n$ is obtained by blowing up $\mathbb{CP}^2$ at $n$ generic points, and a \textit{diagonal} del-Pezzo is a divisor on the threefold $X$ which is a $dP_n$ surface and can be arbitrarily contractible to a point. Therefore, the intersection polynomial of a K3-fibred CY threefold with $h^{1,1} = 3$ and having a diagonal del-Pezzo divisor can be given as:}
\bea
I_3 = k_{122} \, D_1\, D_2^2 + k_{sss} \, D_s^3, 
\eea 
where $D_s$ corresponds to the diagonal del-Pezzo divisor which does not intersect with $D_1 =$ K3 and $D_2$ in the given divisor basis ${\cal B} = \{D_1, D_2, D_s\}$. It is easy to see that such an intersection polynomial leads to the so-called weak Swiss-cheese volume form of the type given in Eq.~(\ref{eq:K3-volumeform}).

Using the CY threefolds arising from the reflexive polytopes of the KS database it has been found that \cite{Cicoli:2018tcq,Cicoli:2021dhg,Shukla:2022dhz}:
\begin{itemize}

\item
For $h^{1,1} = 2$, there are 39 distinct CY geometries, 22 of which have a volume form of the strong Swiss-cheese type with a diagonal del-Pezzo divisor, while there are 10 K3-fibred CY geometries. However, as it is obvious, there are no K3-fibred CY threefold with additional diagonal del-Pezzo divisor. For this reason we have to minimally look for examples with $h^{1,1} = 3$.

\item
For $h^{1,1} = 3$, there are 305 distinct favourable CY geometries, 132 of which have at least one diagonal del-Pezzo divisor, and therefore are a priory suitable for realising LVS. In addition, there are 136 K3-fibred CY geometries, but only 43 of those have, in addition, a diagonal del-Pezzo divisor to support LVS, i.e. a volume form of the type given in Eq.~(\ref{eq:K3-volumeform}). 

\item
For $h^{1,1} = 4$, there are 2000 distinct favourable CY geometries, 750 of which have at least one diagonal del-Pezzo divisor for realising LVS. In addition, there are 865 K3-fibred CY geometries, but only 171 of those have, in addition, at least a diagonal del-Pezzo divisor to support LVS. 

\item
For $h^{1,1} = 5$, there are 13494 distinct favourable CY geometries, 4104 of which have at least one diagonal del-Pezzo divisor for realising LVS. In addition, there are 5970 K3-fibred CY geometries, but only 951 of those have, in addition, a diagonal del-Pezzo divisor to support LVS. 

\end{itemize}
\noindent

\subsection{Searching the K3-fibred models for \texorpdfstring{\antiD}{anti-D3} uplifting}
\label{sec:search}

The purpose of this section is to look for K3-fibred CY threefolds which are also suitable for the realisation of the uplifting scenario based on the presence of an \antiD- brane at the tip of a warped throat, as reviewed in Section \ref{sec:dSuplift}.

In order to do so, we have scanned over all the CY threefolds constructed as hypersurfaces in toric varieties with $h^{1,1}=3,4$ 
classified in the database \cite{CYdatabase}, selecting only the geometries with the following features: 
\begin{enumerate}

\item 
\textit{LVS moduli stabilisation :} A suitable geometry has a negative Euler characteristic (corresponding to $h^{1,2}>h^{1,1}>1$) and at least a rigid toric divisor, supporting non-perturbative effects.
	 In particular we select only geometries presenting at least one `diagonal' del-Pezzo divisor.
	
\item 
\textit{K3-fibration :} The CY threefold is a K3-fibred CY, in addition to having at least one diagonal del-Pezzo divisor. 

\item 
\textit{dS uplift :} To find an appropriate reflection involution $(z_i\rightarrow-z_i)$ such that:

\begin{itemize}

\item 
There are at least two $O3$-planes that come on top of each other under a certain complex structure deformation, so that it is possible to reproduce the required configuration for the \antiD uplift. 

\item 
The overall negative contribution to the D3-charge, coming from $O3$-planes, $O7$-planes and D7-branes is large, so that it is easier to satisfy the constraints \eqref{eq:fluxCons} within the limits of the tadpole cancellation condition. 
This happens in particular when the fixed point locus under the involution includes $O7$-planes wrapping divisors with a large Euler characteristic. 
\end{itemize}
\end{enumerate}

\noindent
The combination of these requirements appears to be very constraining, at least as far as one considers a small number of \K moduli.
First, the requirement of having both a `diagonal' dP and a K3 imposes a strong constraint to the intersection tensor.
The reason behind this obstruction can be anticipated by noting that for a favorable CY threefold with $h^{1,1} = 3$ obtained from a four-dimensional reflexive polytope, there are 7 GLSM charge vectors corresponding to 7 so-called coordinate divisors.
Demanding that one divisor is a K3 and another is a diagonal del-Pezzo leaves very little room for the third divisor in the two-form basis, due to the fact that the K3 appears only linearly in the intersection polynomial and the del-Pezzo, being diagonal, does not intersect with anything else in a suitable choice of divisor basis. 
However, these requirements are not too strong, as one can still get 43 K3-fibred CYs with one diagonal del-Pezzo divisor for $h^{1,1} = 3$ \cite{Cicoli:2018tcq}, and $171$ K3-fibred CYs with at least one diagonal del-Pezzo divisor for $h^{1,1}=4$ \cite{Shukla:2022dhz}. So practically we need to analyse 214 CY geometries for the \textit{dS uplift} conditions listed above. 

The main obstacle in our goal arises instead from demanding the presence of at least two $O3$-planes in the desired configuration for an involution which at the same time can produce a significantly large (in absolute value) contribution to the $D3$-charge tadpole, e.g. via the introduction of a Whitney brane. For the minimal setting of K3-fibred CYs with a diagonal del-Pezzo, corresponding to $h^{1,1} = 3$, this challenge can also be anticipated by the fact that out of the 7 coordinate divisors\footnote{Recall that the favourable CY geometries are defined through the four-dimensional reflexive polytopes of the KS database corresponds to four-dimensional Ambient space, and hence results in a total of 7 toric divisors for $h^{1,1} = 3$ and 8 of them for $h^{1,1} = 4$.}, four would be already unsuitable for our purpose; one corresponding to the K3 and the three other rigid coordinate divisors arising with GLSM charge vectors $\{1,0,0\}, \{0,1,0\}$ and $\{0,0,1\}$, which usually do not possess large $\chi$ nor large cubic intersections\footnote{In fact, for the K3 divisors the cubic self intersection vanishes, while for dP$_n$ surfaces it is given as $\kappa_{sss} = 9-n$, see e.g. \cite{Cicoli:2016xae}. However, as an exception a rigid divisor with large $\chi\ (\chi=111)$ can appear for $h^{1,1}=3$ \cite{Shukla:2022dhz}.}.
 In addition to this, typically the coordinate with the largest weights (associated to the largest Euler characteristic and therefore to the largest D3-charge contribution), does not admit the needed $O3$-planes setup. 
In our detailed scan of K3-fibred CY orientifolds with $2<h^{1,1}<5$, we observe that on most of the cases one can manage to have either large $D3$ tadpole charge or $O3$-planes that can collapse on top of each other (and not both), for a given choice of involution. In fact, for a given CY threefold it is very common and effortless to find an involution resulting in large $D3$ tadpole charge along with another involution resulting in the suitable $O3$-plane configuration, but the main challenge, at least for the small values of $h^{1,1}$ under consideration, is to find an involution which does both these jobs. Nevertheless there are indeed a few examples in which one can manage to have $|Q_{D3}^{O3/O7/D7}| \geq 100$ along with the needed $O3$-planes. 

To be more concrete, after having analysed all the models with $h^{1,1}=3,4$ satisfying the requirements $(1)$ and $(2)$, we noticed that:
\begin{itemize}
	\item The maximal Euler characteristic that we find for a given divisor is $\chi(D)=232$ for $h^{1,1}=3$ and $\chi(D)=435$ for $h^{1,1}=4$. 
However none of the models (geometry + involution) with an $O7$-plane wrapping these divisors includes also $O3$-planes in the desired configuration.
	\item Restricting our search to models with an involution including $O3$-planes, we find that the maximal Euler characteristic for $h^{1,1}=3$ is $\chi(D)=54$, which slightly increases to $\chi(D)=66$ for $h^{1,1}=4$. 
\end{itemize}  
Computing the total negative D3-charge contribution for the selection of models which appear more promising from the point of view of the Euler characteristic of divisors wrapped by $O7$-planes, we find that the maximal (in absolute value) possible value for this quantity, among models with $3$ or $4$ \K moduli is\footnote{Assuming that all the stacks of branes wrapping rigid divisors are fluxless.}
\[-Q_{D3}^{O3/O7/D7}=127\:,\]
corresponding to a model with $h^{1,1}=4$ and an $O7$-plane wrapping a divisor with $\chi(D)=65$.
In the next section, we will analyse this model in detail. 

\subsection{Sub-leading corrections to fix the LVS flat directions}
Generically, the LVS scheme allows to stabilise only a few \K moduli: for example in the minimal models with two K\"ahler moduli, one can have one `large' cycle with volume denoted as $\tau_b$ and one `small' cycle with volume $\tau_s$ leading to the so-called strong Swiss-cheese volume form as given in Eq.~(\ref{eq:swiss-cheese-volume}). In this minimal model using a combination of two subleading contributions, one appearing as $\alpha^\prime$ correction to the K\"ahler potential and the other being non-perturbative effects to the superpotential, one can fix both of the moduli. In fact as we have explicitly revisited in the $h^{1,1} = 2$ case, this mechanism helps in fixing the overall volume ${\cal V}$ of the CY threefold and the $\tau_s$ modulus. However, for models of larger $h^{1,1}$ which result in several K\"ahler moduli, the standard LVS scheme can only stabilise the overall volume ${\cal V}$ and the four-cycle volume moduli $\tau_{s_i}$ which appear in the non-perturbative superpotential via wrapping E3-instantons or D7-stacks inducing gaugino condensation effects via rigid divisors. However, the conditions for generating non-perturbative effects allowing to stabilise the `small' cycles can be very constraining. In order to fix the remaining moduli, therefore, one needs to include sub-leading corrections to the scalar potential: since we are considering a compact CY threefold whose volume $\vol$ has been fixed in LVS at the sub-leading level with respect to the complex structure moduli and axio-dilaton, we expect these corrections to be further suppressed so that they do not spoil the LVS minimum.  

\subsubsection{String-loop corrections}\label{sec:stringLoopCorr}
String-loop corrections to the \K potential have been computed for toroidal models through various routes \cite{Berg:2004ek, Berg:2005ja, Berg:2005yu, Berg:2007wt}, and have been subsequently conjectured for generic CY orientifolds \cite{Cicoli:2007xp}.
We consider in particular two types of string loop contributions: the so-called \textit{Winding} loop corrections arise whenever two divisors wrapped by $O$-planes or $D$-branes intersect each other, admitting a non-contractible one-cycle at the intersection locus; in this case, there is an exchange of closed strings winding the non-contractible cycle; the \textit{Kaluza-Klein} (KK) corrections, come instead from the exchange of KK modes between non-intersecting $D$-branes/$O$-planes.
In \cite{vonGersdorff:2005bf,Cicoli:2007xp}, it was shown that the scalar potential is protected against the leading order pieces of such corrections due to the so-called `extended' no-scale structure  so that they appear in the scalar potential at subleading order in the $\frac{1}{\vol}$ expansion.

The KK and winding loop corrections have been conjectured \cite{Berg:2004ek, Berg:2005ja, Berg:2005yu, Berg:2007wt, Cicoli:2007xp} to take the following form in Einstein frame: 
\beq
\label{eq:KgsE}
	\delta {\cal K}_{g_s}^{\rm KK} = g_s \sum_\alpha \frac{C_\alpha^{\rm KK} \, t^\alpha_\perp}{\cal V} \,, \qquad \delta {\cal K}_{g_s}^{\rm W} =  \sum_\alpha \frac{C_\alpha^W}{{\cal V}\, t^\alpha_\cap}\,,
\eeq
where $C_{\alpha}^{KK}$ and $C_{\alpha}^W$ are functions generically depending on the complex structure and open string moduli, while the $2$-cycle volume moduli $t^{\alpha}_\perp$ denote the transverse distance among stacks of non-intersecting $D7$-branes and $O7$-planes.
The $2$-cycle $t^{\alpha}_\cap$, instead, is the volume of the curve sitting at the intersection loci of the intersecting stacks of D-branes/O-planes.  
Some concrete realisations of these ans\"{a}tze have been presented in explicit CY orientifold settings in~\cite{Cicoli:2016xae, Cicoli:2017axo}.
The scalar potential contributions arising from the corrections in \eqref{eq:KgsE} are given by~\cite{Cicoli:2007xp}:
\bea
\label{eq:VgsKK-W}
& & \delta V_{g_s}^{\rm KK} =  \frac{g_s^3}{2} \frac{|W_0|^2}{\vol^2} \sum_{\alpha\beta} C_\alpha^{\rm KK} C_\beta^{\rm KK} \, {\cal K}_{\alpha\beta}^0 \,, \\
& &\delta V_{g_s}^{\rm W} = -g_s\, \frac{|W_0|^2}{\vol^2} \, \delta {\cal K}_{g_s}^{\rm W} = -g_s\, \frac{|W_0|^2}{\vol^3} \, \sum_\alpha \frac{C_\alpha^W}{t^\alpha_\cap}\,. \nonumber
\eea
Here, ${\cal K}_{\alpha\beta}^0$ is the tree-level K\"ahler metric:
\bea
\label{eq:Kij-tree}
& & {\cal K}_{\alpha\beta}^0 = \frac{1}{16\, {\cal V}^2}\left(2\,t^\alpha\, t^\beta - 4 {\cal V}\, k^{\alpha\beta} \right)\:,
\eea
with $k^{\alpha\beta}=(k_{\alpha\beta})^{-1}=(k_{\alpha\beta\gamma}t^\gamma)^{-1}$.

We also mention that in a recent revisit of the string-loop corrections in \cite{Gao:2022uop}, it has been found that the Winding type effects can appear more generically as to what is expected from the original proposals of \cite{Berg:2004ek, Berg:2005ja, Berg:2005yu, Berg:2007wt}. This is also consistent with an earlier field theoretic analysis performed in \cite{vonGersdorff:2005bf}. In addition to that, apart from the KK-type and Winding type string loop effects discussed so far, there can be additional one-loop effects appearing as logarithmic corrections in the K\"ahler potential \cite{Antoniadis:2018hqy, Antoniadis:2019rkh}. Such terms can arise from specific configurations of $D7$ brane stacks and from a four-dimensional Einstein-Hilbert term localised within the six-dimensional internal space, originally being generated from higher derivative terms in the ten-dimensional string effective action. However, given the specific need to have (a minimum of) 3 stacks of $D7$-branes with appropriate intersection loci, we do not expect such terms to get generically induced, without the need of making any specific flux engineering on appropriate cycles wrapping the brane stacks. For these reasons we will not consider these terms in the moduli stabilisation analysis of the current work.

\subsubsection{Higher derivative \texorpdfstring{$F^4$}{F4} corrections}
In addition to the string loop corrections, we consider also the higher derivative correction appearing at $\mathcal{O}(F^4)$ in the scalar potential \cite{Ciupke:2015msa}. This correction is argued to be generic for a given CY orientifold compactification and takes the simple form: 
\beq
\label{eq:VF4}
	 V_{F^4} = - \frac{g_s^2}{4} \frac{\lambda\,|W_0|^4}{g_s^{3/2} {\cal V}^4} \Pi_i \, t^i,
\eeq
where the $t^i$'s are the volumes of the $2$-cycles for the generic CY manifold $X_3$ and the $\Pi_i$ are topological numbers, also called \textit{second Chern numbers}, defined as:
\beq
	\Pi_i =\int_{D_i}c_2(X_3)\:.
\eeq
In a recent work \cite{Shukla:2022dhz}, all the divisors of distinct topologies along with their respective values of $\Pi_i$ corresponding to the CY threefolds with $1 \leq h^{1,1} \leq 5$ arising from the KS database have been presented. We also note that, $\lambda$ is an unknown combinatorial factor whose value is expected to be between $10^{-2}$ and $10^{-3}$ \cite{Grimm:2017okk}. However, this estimated value of $\lambda$ corresponds to a single K\"ahler modulus model based on CY compactifications with $h^{1,1}(CY) =1$.


\section{An explicit model}
\label{sec:explicit_model}

In this section we present an explicit example of a  CY threefold with $h^{1,1} = 4$ which has a K3-fibration structure and possesses a diagonal del-Pezzo divisor leading to the CY volume form given as below,
\bea
\label{eq:vol-type}
& & {\cal V} = f_1(t_i^\prime) + \, t_{\rm K3} \, f_2(t_i^\prime) + a t_s^3\:,
\eea
where $a$ is a constant; $t_s$ and $t_{\rm K3}$ correspond to the size of the $2$-cycles dual to the del-Pezzo and the K3 divisor respectively, and $f_1(t_i^\prime), f_2(t_i^\prime)$ are homogeneous polynomials of degree three and two respectively  in the remaining $2$-cycle volumes $t_i^\prime$.

As mentioned in the previous section, we choose, in particular, the model (geometry $+$ involution) corresponding to the largest negative D3-charge contribution.

\subsection{Geometric data}

The CY threefold of our interest corresponds to the polytope ID\#1271 (Triangulation\#1) in the CY database of \cite{Altman:2014bfa} and it is defined by the following toric data:
\begin{center}
	\begin{tabular}{|c c c c c c c c|c|}
		\hline
		$x_1$  & $x_2$  & $x_3$  & $x_4$  & $x_5$ & $x_6$  & $x_7$   & $x_8$ & $D_H$  \\
		\hline
		0 & 0 & 0 & 0 & 0 & 1 & 2 & 1 &4 \\
		0 & 0 & 0 & 1 & 1 & 1 & 3 & 0 & 6\\
		1 & 0 & 1 & 2 & 0 & 3 & 7 & 0 & 14\\
		1 & 1 & 0 & 0 & 0 & 2 & 4 & 0 & 8\\
		\hline
		\hline
		 SD1 & NdP$_9$ & NdP$_{10}$ & K3& ${\mathbb P}^1\times{\mathbb P}^1$ & $D_6$ &$D_7$ & dP$_1$ & Topology \\
		 \hline		 
	\end{tabular}
\end{center}
with SR-ideal
\beq\label{eq:ID1271SR}
	 {\rm SR} =  \{x_1 x_2, x_2 x_5, x_5 x_8, x_6 x_8, x_1 x_3 x_6, x_3 x_4 x_7, x_4 x_5 x_7\} 
\eeq
and second Chern numbers:
\beq\label{eq:ID1271_2ndChNum}
	\begin{split}
	 	\Pi_i=\int_{D_i}c_2(X) = \lbrace 26, 12, 14, 24, -4, 58, 126, -4 \rbrace\:.
	\end{split}
\eeq
The Hodge numbers for the CY threefold are $(h^{2,1}, h^{1,1}) = (102, 4)$, and hence the Euler characteristic is given as $\chi(X_3)=-196$.

The analysis of the divisor topologies, performed by means of {\tt cohomCalg} \cite{Blumenhagen:2010pv, Blumenhagen:2011xn} shows that $D_4$ is a K3 surface. 
Moreover, the divisors $D_5$ and $D_8$ both have the Hodge numbers of a del-Pezzo divisor dP$_1$; however, after applying the criteria of \cite{Cicoli:2021dhg} it turns out that $D_5$ is a diagonal  ${\mathbb P}^1\times{\mathbb P}^1$ surface, while $D_8$ is a non diagonal dP$_1$.
Further, the divisors $D_2$ and $D_3$ correspond to the so-called `rigid-but-not-del-Pezzo' divisors,  NdP$_9$ and NdP$_{10}$ respectively, whose names are derived by the fact that their Hodge diamond resembles the one of dP surfaces\footnote{A dP$_n$ divisor has Hodge numbers $h^{\bullet}(dP_n)\equiv\lbrace h^{0,0}(dP_n), h^{0,1}(dP_n), h^{0,2}(dP_n), h^{1,1}(dP_n)\rbrace=\lbrace1,0,0,n+1\rbrace$, with $n=0,...,8$; a NdP$_n$ divisor has the same Hodge numbers but with $n>8$.}. 
Finally, $D_1$ is a so-called `special deformation' divisor SD1 and $D_6, D_7$ are deformation divisors with the Hodge diamonds reported below:
\bea
{\rm SD1}  :  \begin{tabular}{ccccc}
    & & 1 & & \\
   & 0 & & 0 & \\
  1 & & 21 & & 1 \\
   & 0 & & 0 & \\
    & & 1 & & \\
  \end{tabular},
\eea
\bea
  D_6 : \begin{tabular}{ccccc}
    & & 1 & & \\
   & 0 & & 0 & \\
  5 & & 53 & & 5 \\
   & 0 & & 0 & \\
    & & 1 & & \\
  \end{tabular}, \qquad \quad D_7 : \begin{tabular}{ccccc}
    & & 1 & & \\
   & 0 & & 0 & \\
  24 & & 163 & & 24 \\
   & 0 & & 0 & \\
    & & 1 & & \\
  \end{tabular}\:.
  \nonumber
\eea
In the following, we will work in the basis of smooth divisors:
\beq\label{eq:basis}
	\mathcal{B}=\lbrace D_2, D_4, D_5, D_8 \rbrace\:.
\eeq
Expanding the K\"ahler form over this basis, $J=t_2 D_2+t_4 D_4+ t_5 D_5+t_8 D_8$, we get the intersection polynomial
\beq\label{eq:1271I3}
	I_3 = 8 D_5^3-D_2^2 D_8+2 D_2 D_4 D_8-D_2 D_8^2-4 D_4 D_8^2+8 D_8^3	
\eeq
and subsequently the overall volume of the CY threefold given as:
\beq\label{eq:ID1271vol}
	\begin{split}
	\vol &= \frac{1}{6}\int_{X_3} J^3=\frac{1}{6}k_{ijk}t^i t^j t^k \\
	 &= -\frac{t_2^2\, t_8}{2} -\frac{t_2 t_8^2}{2} +\frac{4 t_8^3}{3} + 2 t_2 t_4 t_8 -2 t_4 t_8^2 +\frac{4 t_5^3}{3}\:.
	 \end{split}
\eeq
We notice that the diagonality of the ${\mathbb P}^1\times{\mathbb P}^1$ divisor $D_5$ and the linearity of the K3 divisor $D_4$ are manifested in the intersection polynomial \eqref{eq:1271I3} as well as in the volume form \eqref{eq:ID1271vol}. 
In particular, referring to Eq. \eqref{eq:vol-type}, we find that our model corresponds to:
\bea
& & f_1 = -\frac{t_2^2\, t_8}{2} -\frac{t_2 t_8^2}{2} +\frac{4 t_8^3}{3}, \quad f_2 =  2 t_2 t_8 -2 t_8^2, \quad a = \frac43\:.
\eea
The four-cycle volumes $\tau_i$ of the basis divisors, can be expressed in terms of the 2-cycles $t_i$ as below ($\tau_i=\frac{\partial\vol}{\partial t_i}$):
\begin{equation}\label{eq:ID1271tau}
	\begin{split}
		\tau_2 &=-t_2\, t_8 + 2\, t_4\, t_8 - \frac{t_8^2}{2}	\:; \qquad \tau_4=2 \,t_2\, t_8 - 2 t_8^2 \:;\\
		\tau_5 &=4t_5^2; \qquad \tau_8=-\frac{t_2^2}{2} + 2\, t_2\, t_4 -  t_2\, t_8 - 4\, t_4\, t_8 + 4 t_8^2\:.
	\end{split}
\end{equation}
However, it is not possible to express the volume form $\vol$ exclusively in terms of the $4$-cycle volumes in a simple form, due to the presence of the non-diagonal dP divisor $D_8$.
Finally,  the \K cone conditions are given as:
\beq\label{eq:ID1271KCC}
	2 t_5 + t_8 > 0\:, \quad -2 t_5 > 0\:, \quad -t_2 + 2 t_4 - t_8 > 0\:, \quad t_2 - 2 t_8 > 0\:.
\eeq

\subsection{Involution, brane-setting and the tadpole charge}

We consider the involution: 
\beq\label{eq:ID1271inv}
	x_6 \rightarrow -x_6\:,
\eeq
for which there are 2 $O7$-planes located at $D_6$ and $D_8$ respectively, a single $O3$-plane at $x_1=x_3=x_7=0$ and two $O3$-planes at $x_2=x_3=x_7=0$.

We have therefore $\sum_i D_{O7_i}^3\equiv D_6^3+D_8^3=15$, from which we can compute the correction \eqref{eq:chiCorr} to the Euler characteristic:
\beq\label{eq:chiCorr}
	\chi(X_3)\rightarrow\chi(X_3)+2\int_{X_3}D_{O7}^3=-166\:.
\eeq
Moreover, after computing $\chi(O_\sigma)=\chi(D_6)+\chi(D_8)+3\chi(O3)=65+4+3=72$, we can use the Lefschetz fixed point theorem to derive the values of $h^{1,2}_+$ (number of abelian bulk vectors) and $h^{1,2}_-$ (number of complex structure deformations of the invariant equation of the CY).
For a CY threefold with $h^{1,1}_-(X)=0$, in particular, the theorem has the simple formulation \cite{Crino:2022zjk}:
\beq
	2(h^{1,1}(X)-h^{1,2}_+(X)+h^{1,2}_-(X)+2)=\chi(O_\sigma)\:,
\eeq
from which we obtain, recalling also that $h^{1,2}=h^{1,2}_+(X)+h^{1,2}_-(X)$, the values $h^{1,2}_+(X)=36;\ h^{1,2}_-(X)=66$.

\paragraph{}

Each $O3$-plane contributes to the total D3-charge with $Q_{D3}^{O3}=-\frac{1}{2}$, while each $O7$-plane has a charge $Q_{D3}^{O7}=-\frac{\chi(D)}{6}$, where $\chi(D)$ is the Euler characteristic of the divisor wrapped by the $O7$-plane.
The total contribution to the D3-charge coming from the Op-planes is therefore 
\bea
& & Q_{Op} = -\frac{1+2}{2} - \frac{65+4}{6} = -13\:.
\eea

In order to cancel the D7-charge of the $O7$-planes wrapping the divisors $D_6$ and $D_8$, we need to introduce suitable configurations of D7-branes in the class: $[D_{D7}]=8[D_{O7}]$. 
Since $D_8$ is a rigid divisor, its D7-charge is cancelled by a stack of four D7-branes (plus four image branes) wrapping the same divisor. 
The D7-charge of the O7-plane wrapping the non-local divisor $D_6$ will instead be compensated by the introduction of a Whitney brane \cite{Collinucci:2008pf,Collinucci:2008sq}.

Further, we recall that a D7-brane wrapping a divisor $D$ supports a gauge invariant flux $\mathcal{F}=F-\iota^*B$, where $F$ is the gauge flux, $\iota^*$ denotes the pull-back on the divisor $D$ and $B$ is the  NSNS $2$-form field.
The gauge flux must be quantised such that the Freed-Witten anomaly is cancelled \cite{Freed:1999vc}, that is:
\beq
	F+\frac{c_1(D)}{2}\in H^2(D,\mathbb{Z}),
\eeq
where $c_1(D)=-\iota^*D$ in a CY.

By choosing a $B$-field $B=\frac{D_8}{2}$, we can take a quantised flux $F$ such that $\mathcal{F}_8=0$. 
In addition to this, we notice that the diagonal ${\mathbb P}^1\times{\mathbb P}^1$ $D_5$ supports the presence of an $E3$-instanton, which generates a non-perturbative effect to the superpotential (as needed for the stabilisation of the corresponding \K modulus via LVS) \textit{provided that} $\mathcal{F}_5=0$.
Since $x_5 x_8\in SR$ \eqref{eq:ID1271SR}, we can choose
\beq
	B=\frac{D_5}{2}+\frac{D_8}{2}\:,
\eeq
therefore ensuring that both $\mathcal{F}_5$ and $\mathcal{F}_8$ vanish. 
In this case, the D7-branes stack on $D_8$ supports a $SO(8)$ pure super Yang-Mills. 
The D3-charge of the $SO(8)$ stack is given by\footnote{The D3-charge for a single D7-brane with $\mathcal{F}\neq0$ reads $Q_{D3}^{D7}=-\frac{\chi(D)}{24}-\frac{1}{2}\int_D\mathcal{F}\wedge\mathcal{F}$.}:
\beq
	Q_{D3}^{D7}=-\frac{(4+4)}{24}\chi(D_8)=-\frac{4}{3}.
\eeq
In order to compute the D3-charge contribution from the Whitney brane wrapping $D_6$, we can instead use the generic formula $Q_{D3}^W(D)=Q_{D3}^{W,\,\rm geom}(D)+Q_{D3}^{W,\,\rm flux}(D)$, with \cite{Crino:2022zjk}
\begin{eqnarray}
\label{eq:D3WD7geom}Q_{D3}^{W,\,\rm geom}(D) &=&  -  \frac{1}{3}\int_{X_3} D\wedge \left( 43 D\wedge D + c_2(X) \right) = -\frac{\chi(4D)}{12} -9\int_{X_3} D^3\:, \\
\label{eq:D3WD7flux}Q_{D3}^{W,\,\rm flux}(D)&=& \int_{X} D \wedge (3D-2F+2B)\wedge (3D + 2F - 2B)\:.
\end{eqnarray}
The flux on the brane is determined by the two-form $F=\sum_i a_i D_i\  (a_i\in \mathbb{Z})$, which must satisfy the condition 
\beq\label{eq:fluxCond}
	-\frac{3D}{2}+B\leq F\leq\frac{3D}{2}+B\:,
\eeq
so that the Whitney brane is not forced to split into a brane/image-brane system.

Since $\chi(4D_6)=680$ and $D_6^3=7$, the geometric D3-charge contribution for the Whitney brane wrapping the divisor $D_6$ turns out to be:
\[Q_{D3}^{W,\,\rm geom}(D)=-\frac{359}{3}\:.\]
As concerns the flux contribution \eqref{eq:D3WD7flux}, we notice that since it is positive we need to minimise it.
Among the possible choices satisfying \eqref{eq:fluxCond}, we select, in the basis \eqref{eq:basis}, $F=-3D_2-2D_4+D_5+3D_8$, for which we have $Q_{D3}^{W,\,\rm flux}=7$ and therefore:
\[Q_{D3}^{W}=-\frac{359}{3}+7=-\frac{338}{3}.\]
In summary, the total D3-charge arising from the Op-planes and the D-branes is:
\beq
	Q_{D3}=-13 -\frac43 - \frac{338}{3} = -127\:.
\eeq

\subsection{O3-planes at the tip of the warped throat}

Let us conclude the analysis of the geometric features of our model, by proving that a highly warped throat is generated in a corner of the complex structure moduli space and that there is a pair of $O3$-planes at the tip of this throat.
This is indeed the required setup for the implementation of the uplift mechanism introduced in Section \ref{sec:dSuplift}. 
In order to do so, we follow \cite{Garcia-Etxebarria:2015lif}\footnote{See also \cite{Crino:2020qwk} for a completely analogous computation, presented in more detail.}.

The equation defining the CY three-fold, once restricted to its invariant version with respect to the involution \eqref{eq:ID1271inv}, has the following structure:
\beq\label{eq:CYeq}
	\begin{split}
		x_7^2&= a \,x_5\, x_6^4\, (x_4+a_1\, x_3^2x_5)-2b \,x_6^2\, x_8^2\,\left[x_1^4\, x_4^2\, x_5^2+x_2\, P_1(x_1,x_2,x_3,x_4,x_5)\right.\\
		&\left.+x_3\, P_2(x_1,x_3, x_4,x_5)\right] + x_2\, P_3(x_1,x_2,x_3,x_4,x_5,x_8)+x_3\, P_4(x_1,x_3,x_4,x_5,x_8)\\
		&+c\, x_1^8\, x_4^3\, x_5^3\, x_8^4=0
	\end{split}
\eeq
where the $P_i(x_j)$ are polynomials in the coordinates $\{x_j\}$ and $a, a_1,b,c\in\mathbb{C}$.
At the locus $x_2=x_3=x_7=0$ of the $2$ $O3$-planes, therefore, it becomes:
\beq\label{eq:O3eq}
	x_4\, x_5\, x_6^4-2b\, x_1^4\, x_4^2\, x_5^2\, x_6^2\, x_8^2+c\, x_1^8\, x_4^3\, x_5^3\, x_8^4=0
\eeq
(where we redefined the constants $b,c$ in order to get rid of $a$).

Looking at the SR-ideal \eqref{eq:ID1271SR}, we notice that the coordinates $\{x_1, x_4,x_5\}$ can never vanish at the locus of these O3-planes.
Fixing them to $1$ (by an appropriate choice of the $\mathbb{C}^*$ projective rescaling parameters), \eqref{eq:O3eq} becomes:
\beq
	x_6^4 - 2 b\,x_6^2\, x_8^2 + c\, x_8^4=0\:.
\eeq
Furthermore, $x_6$ and $x_8$ cannot vanish simultaneously. We can therefore fix $x_8=1$ as well and obtain:
\beq\label{eq:O3gfEq}
	x_6^4 - 2 b\,x_6^2 + c =0\:.
\eeq
This equation is completely analogous to the one obtained for the explicit model of \cite{Crino:2020qwk}, hence also in this case we can redefine $c\equiv b^2-\delta$, so that when $\delta=0$ the two O3-planes, sitting at the two zeroes of \eqref{eq:O3gfEq}\footnote{Calling $\gamma_i$ the zeroes of the quadratic equation, it is straightforward to see that the equation is solved by $x_6^2=\gamma_i$. Moreover, the solutions $x_6=\pm\gamma_i$ are identified by the $\mathbb{Z}_2$ orbifold action.} go on top of each other at the point $x_6^2-b\, x_8^2=0$, while they are very close to each other when $\delta$ is small.

By analysing the neighbourhood of the point $x_2=x_3=x_7=(x_6^2-b\, x_8^2)^2+\delta=0$, we can see that the equation of the CY \eqref{eq:CYeq} in an ambient space $\mathbb{C}^4/\mathbb{Z}_2$ is precisely the one of a deformed conifold:
\beq
	-x_7^2+x_2^2+x_3^2+(x_6^2-b)^2+...=\delta
\eeq
which, in the limit $\delta\rightarrow0$ develops a conifold singularity at the point  $x_2=x_3=x_7=(x_6^2-b)^2=0$.
The parameter $\delta$, therefore, parametrises the size of the contracting $S^3$ at the tip of the throat.

Finally, by carefully inspecting the gauge fixing introduced before, we can see that in the local patch under consideration, the involution reproduces exactly the geometric action required for the retrofitting of a nilpotent Goldstino sector \cite{Garcia-Etxebarria:2015lif}, that is:
\beq
	x_7\rightarrow-x_7, \quad x_2\rightarrow-x_2, \quad x_3\rightarrow-x_3\:.
\eeq
We conclude that our set up is suitable for a configuration in which we add a D3 brane at one of the $O3^-$ points of the contracting $S^3$ and a \antiD at the other $O3^-$ point.
This ensures, as anticipated,  that the D3/\antiD branes do not contribute to the total D3-charge and that, if the complex structure moduli are fixed so that the $S^3$ has finite size ($\delta\neq0$), there is no perturbative decay channel between them, as they are stuck at the $O3$-planes loci \cite{Garcia-Etxebarria:2015lif}.

\section{Moduli stabilisation and dS realisation}
\label{sec:modStab}

\subsection{Collecting the scalar potential pieces from various sources}

In our concrete global construction, there can be several scalar potential contributions sourced from various (non-)perturbative effects which we use for moduli stabilisation. 
In order to have an analytical control over our results, we stabilise the moduli in the following three steps:
\begin{itemize}
	\item First, we consider the complex structure moduli (with the exception of the complex structure modulus $\zeta$, parametrising the highly warped throat) and the axio-dilaton to be stabilised at leading order, which corresponds to the scalar potential pieces scaled as ${\cal V}^{-2}$ in the inverse volume expansion. 

	\item The overall volume modulus ${\cal V}$ and the `small' divisor volume $\tau_s = \tau_5$ corresponding to the diagonal ${\mathbb P}^1 \times {\mathbb P}^1$ divisor are stabilised in the standard LVS scheme, using the non-perturbative effect generated by the $E3$-instanton on the divisor $D_5$.
In principle one should also include the non-perturbative effect coming from the $SO(8)$ stack of D7-branes wrapping $D_8$, which, being fluxless, supports gaugino condensation.
	However, given that the divisor $D_8$ is non-diagonal, it is plausible to assume that the intersection with the other divisors produces unwanted zero modes that might kill the non-perturbative effect. 
	In addition to this, from the KCC \eqref{eq:ID1271KCC}, we can see that $D_8$ is expected to be a `large' cycle due to its non-local nature; as a consequence, even if the non-perturbative effect is non-vanishing, it is reasonable to assume it to be highly sub-dominant and therefore useful only for the purpose of fixing the non-local axionic moduli corresponding to the complexified four-cycle volume modulus $T_8$. We will see, indeed, that at least for our explicit choice of the parameters, the $2$-cycle $t_8$ turns out to be stabilised at a large enough value to fully justify (a posteriori) this assumption. 

In this step, we also include the uplifting term and we consider explicitly the stabilisation of the complex structure modulus $Z$ parametrising the throat. 

\item The remaining two \K moduli are fixed by sub-leading corrections arising from string loop effects together with the higher derivative $F^4$ corrections (see Section \ref{sec:subLeadCorr}), which are beyond the reach of the two-derivative approximation captured by the K\"ahler and superpotential description.

\end{itemize}
\noindent

\subsubsection*{LVS potential with \texorpdfstring{\antiD}{anti-D3} uplift:}

Apart from the BBHL's $(\alpha^\prime)^3$ corrections \cite{Becker:2002nn}, the presence of a diagonal ${\mathbb P}^1 \times {\mathbb P}^1$ divisor facilitates, as mentioned above, an $E3$-instanton contribution to the superpotential wrapping the divisor $D_5$:
\bea
& & W_{np} = A_5 \, e^{-a_5\, T_5},
\eea
where $a_5 = 2\,\pi$ and $A_5$ can be considered as a parameter after fixing the complex structure moduli and the axio-dilaton by the leading order effects. 
Computing the scalar potential as in \cite{Crino:2020qwk}, we find: 
\beq\label{eq:V-LVS}
	\begin{split}
		 V_{\rm LVS+up} &= \frac{8 g_s a_5^2 A_5^2 e^{-2 a_5 \tau_5} \sqrt{\tau_5}}{\vol} +\frac{2 g_s a_5 A_5 \, |W_0| \tau_5 \, e^{-a_5 \tau_5} \cos(a_5 \theta_5 + \phi)}{\vol^2} +\frac{3 |W_0|^2 \xi}{8\sqrt{g_s}\vol^3}\\
		 &+\frac{c'' \zeta^{4/3}}{2 g_s M^2 \pi \vol^{4/3}}+\frac{\zeta^{4/3}}{2c' M^2 \vol^{4/3}}\left(\frac{K^2}{g_s^2} + \frac{M^2 \sigma^2}{4 \pi^2} + \frac{M K}{g_s \pi} \log\zeta + \frac{M^2}{4 \pi^2} \log^2\zeta\right)\:.
	\end{split}
\eeq
Stabilising as usual the complex structure modulus $\zeta$ and the two axions $(\theta_5,\sigma)$, we get (see also Eq. \eqref{eq:zMin}):
\bea
& & \theta_{5}=\frac{\pi-\phi}{a_5}; \quad \sigma_0=0
\eea
and
\bea
\label{eq:zeta0}
& & \zeta_0 = e^{-\frac{2\pi K}{g_s M}-\frac{3}{4}+\sqrt{\frac{9}{16}-\frac{4\pi}{g_s M^2}c' c''}}.
\eea
These values can be substituted in Eq \eqref{eq:V-LVS}, obtaining:
\beq
	V_{\rm LVS+up} = \frac{8g_s a_5^2 A_5^2 e^{-2 a_5 \tau_5} \sqrt{\tau_5}}{ \vol}-\frac{2 g_s a_5 A_5 \, |W_0| \tau_5\, e^{-a_5 \tau_5}}{\vol^2} +\frac{3g_s |W_0|^2 \hat\xi}{8 \vol^3}+\frac{q_0 \zeta_0^{4/3}}{\vol^{4/3}}\:,
\eeq
where $q_0$ was defined in Eq. \eqref{eq:q0}.

\subsubsection*{String-loop effects:}

Let us now compute the sub-leading corrections coming from string loop effects as described in Section \ref{sec:stringLoopCorr}. 
Since the only divisors wrapped by D7-branes/O7-planes are $D_6$ and $D_8$, which do not intersect each other, no winding correction à la \cite{Berg:2004ek, Berg:2005ja, Berg:2005yu, Berg:2007wt} arises in the considered setup. Using the field theoretic arguments, it has been claimed in \cite{Gao:2022uop} that the winding loop corrections can be more generic than what was considered before. However, given that there are still some implicitness (if not freedom) available in the complex structure moduli dependent coefficients, one may be justified in neglecting them for the current purpose.

On the other side, the presence of $O3$-planes implies that we can generically expect KK string loop corrections, parametrised by the transverse distance between these planes and the divisors wrapping the O7/D7. 
Since we do not know the explicit form of the CY metric, we consider the transverse distance $t^\alpha_\perp$ as a linear combination of all the four two-cycle volume moduli, $t^\alpha_\perp = p_\alpha\, t^\alpha$, where the coefficient $p_\alpha$ can be absorbed in the $C_\alpha^{\rm KK}$ and $C_\beta^{\rm KK}$ (notice that since we are assuming the complex structure moduli to be stabilised in a previous step, these two functions can be now considered as constant parameters). 

The contribution to the scalar potential reads therefore: 
\beq
\label{eq:Vgs-globalmodel}
	V_{g_s} \equiv V_{g_s}^{\rm KK} = \frac{g_s^3}{2} \frac{|W_0|^2}{16\,\vol^4} \sum_{\alpha,\beta} C_\alpha^{\rm KK} C_\beta^{\rm KK} \left(2\,t^\alpha\, t^\beta - 4 {\cal V}\, k^{\alpha\beta} \right)\:,
\eeq
where the matrix $k^{\alpha\beta}$ is the inverse of:
\beq\label{eq:k-alpha-beta}
	k_{\alpha\beta} = \left(
	\begin{array}{cccc}
		-t_8 & 2 t_8 & 0 & -t_2 +2 t_4-t_8 \\
		2 t_8 & 0 & 0 & 2 t_2-4 t_8 \\
		0 & 0 & 8 t_5 & 0 \\
		-t_2+2 t_4- t_8 & 2 t_2-4 t_8 & 0 & -t_2-4 t_4+8 t_8 \\
	\end{array}
	\right)\:.
\eeq

\subsubsection*{Higher derivative \texorpdfstring{$F^4$}{F4} corrections:}

As concerns the higher derivative correction, the corresponding contribution is simply given by Eq. \eqref{eq:VF4}, evaluated in terms of the chosen basis of $2$-cycles, with the $\Pi_i$ reported in \eqref{eq:ID1271_2ndChNum}:
\beq
\label{eq:F4-term-globalmodel}
	 V_{F^4} = -\frac{g_s^2}{4} \frac{\lambda\,|W_0|^4}{g_s^{3/2} {\cal V}^4} \left(12 \, t_2 + 24\, t_4 - 4 \,t_5 -4 \,t_8\right)\:.
\eeq

It could be worth to mention that in both the above corrections, namely those arising from string-loops and $F^4$ effects, there is some implicitness in the coefficients in the sense of their form being not known explicitly. For string-loop effects this depends on the complex structure moduli while in the case of $F^4$ it may be a purely combinatorial constant. However the point which we want to emphasise here is the fact that enforcing (some of) them to always take ${\cal O}(1)$ values may be an overestimation and values of order $10^{-2}$-$10^{-1}$ could still be (naturally) realised while considering all the moduli, including the complex structure and the dilaton, in the dynamics leading to moduli stabilisation.

\subsubsection*{Total effective scalar potential for the volume moduli:}

To summarise, after the minimisation of the axionic moduli $\theta_5, \sigma$ and the throat complex structure modulus $\zeta$, we are left with the following scalar potential for the \K moduli:
\bea
\label{eq:Vmain}
& & \hskip-0.5cm V = V_{\rm LVS+up} + V_{g_s} + V_{F^4} \\
& & = \frac{8g_s a_5^2 A_5^2 e^{-2 a_5 \tau_5} \sqrt{\tau_5}}{ \vol}-\frac{2 g_s a_5 A_5 e^{-a_5 \tau_5} |W_0| \tau_5}{\vol^2} +\frac{3g_s |W_0|^2 \hat\xi}{8 \vol^3}+\frac{q_0 \zeta_0^{4/3}}{\vol^{4/3}} \nonumber\\
& & + \frac{g_s^3}{2} \frac{|W_0|^2}{16\,\vol^4} \sum_{\alpha,\beta} C_\alpha^{\rm KK} C_\beta^{\rm KK} \left(2\,t^\alpha\, t^\beta - 4 {\cal V}\, k^{\alpha\beta} \right) \nonumber\\
& & -\frac{g_s^2}{4} \frac{\lambda\,|W_0|^4}{g_s^{3/2} {\cal V}^4} \left(12 \, t_2 + 24\, t_4 - 4 \,t_5 -4 \,t_8\right)\:, \nonumber
\eea
where $\zeta\equiv\zeta_0$ and $q_0$ are defined in Eq. \eqref{eq:zeta0} and Eq. \eqref{eq:q0} respectively, while $k^{\alpha\beta}$ is the inverse of the matrix defined in Eq. (\ref{eq:k-alpha-beta}). 


\subsection{Numerical moduli stabilisation}

In order to minimise the scalar potential \eqref{eq:Vmain}, we consider the following steps:
\begin{itemize}
	\item Taking into consideration only the leading order term $V_{\rm LVS+up}$, we stabilise the overall volume of the CY $\vol$ and the `small' cycle $\tau_5$. In order to do so, we follow the same strategy presented in \cite{Crino:2020qwk} and summarised at the beginning of Section \ref{sec:LVSstab}, which also includes the choice of the values of the parameters $W_0, g_s$ as well as the fluxes $M,K$.
	\item After having fixed $\vol$ and $\tau_5$, we include also the sub-leading corrections which we use to stabilise the remaining two moduli. Within this step, we also fix the values of the parameters $C_\alpha^{KK}$ and $\lambda$.
\end{itemize}
In order to perform the above two steps, we use simplicial homology global optimisation algorithm~\cite{Endres2018} implemented in the scientific computing tool \texttt{SciPy}~\cite{2020SciPy-NMeth}. 
The optimisation is constrained based on the physical consistency conditions prescribed in Section~\ref{sec:list-of-requirements}. 
The automatic differentiation based package \texttt{JAX}~\cite{jax2018github} is used for computing the gradients and Hessians of the scalar potential. 
The relatively simple scalar potential \eqref{eq:Vmain}, for which we still have a certain amount of analytic control, is a very good environment for the study of these algorithms and for the test of their application to this kind of setups. The final goal is to expand their use to more involved and generic scalar potentials, in the future.

\subsubsection{Stabilisation of the overall volume and the small cycle (LVS)}\label{sec:LVSstab}

Let us consider the term $V_{\rm LVS+up}$, defined in Eq. \eqref{eq:V-LVS}.
Since we are considering, as source of non-perturbative effect, the E3-instanton on the divisor $D_5$, we have $a_5=2\pi$.
The complex-structure depending parameter $A_5$ is instead fixed to $1$ as usual.  

The strategy followed to find solutions for the LVS scalar potential is composed by the following steps:
\begin{enumerate}
	\item We scan over values of $|W_0|\in [1,30]$ \footnote{We analysed also a small selection of non-integer values in the range $0<|W_0|<1$.}, with step $1$, and $g_s\in [0.01,0.55]$, with step $0.001$.
	\item As derived in \cite{Crino:2020qwk}, the requirement of having a dS minimum for the scalar potential imposes a strong constraint on the possible values of the warp factor $\rho$, which turns out to be\footnote{Notice that there is a difference of a factor of $1/2$ with respect to \cite{Crino:2020qwk}, due to the fact that here we are using a slightly different notation for $\xi$.}
	\beq\label{eq:rhoMin}
		\rho\simeq\alpha\frac{27g_s|W_0|^2\kappa_5\,\tau_5^{1/2}}{40\, a_5 \vol^{5/3}}\:,
	\eeq
	with $\alpha\in]1,\frac{9}{4}[$.
	As a second step, therefore, we compute the lower and upper bounds $\rho_{\rm min}$ (corresponding to $\alpha=1$) and  $\rho_{\rm max}\ (\alpha=\frac{9}{4})$, for each combination of values $(|W_0|, g_s)$.
	\item Scanning over integer values of $M$ and $K$ such that 
	\begin{itemize}
		\item $MK<|Q_{D3}^{O3/O7/D7}|=127\:;$
		\item The requirement \eqref{eq:fluxCons} is satisfied;
	\end{itemize}		
	 we compute the value of the warp factor at the minimum: $\rho=q_0\zeta^{4/3}$, with $q_0$ and $\zeta$ defined in \eqref{eq:q0} and \eqref{eq:zeta0} respectively.
	 We select only those combinations of $M,K$ such that $\rho_{\rm min}<\rho<\rho_{\rm up}$.
	\item We check each of the solutions found from the previous step in order to verify whether the requirements of Section \ref{sec:list-of-requirements} are satisfied. 
\end{enumerate}

The procedure outlined before produces a set of viable configurations allowing to consistently stabilise $\vol$ and $\tau_5$ in a de Sitter minimum. 
Among these configurations, we select the one(s) corresponding to the minimal flux contribution to the D3-charge, which turns out to be:
\[MK=100\qquad (M=K=10)\:.\] 

However, a solution with such flux numbers (and in general flux numbers smaller than $127$) requires a relatively large string coupling. 
We choose, as an example\footnote{We find several solutions corresponding to the same flux contribution. All of them have $g_s\sim\frac{1}{2}$.}: 
\[|W_0|=4; \qquad g_s=0.52\:,\]
corresponding to $\zeta=3.44\times10^{-6}$ and $g_s M=5.2$ \footnote{As in \cite{Crino:2020qwk}, we assume $g_sM\gtrsim5$ to be enough in order to trust the KS approximation.}.
The string coupling $g_s$ is therefore quite large, with respect to the values that are commonly taken into consideration for this parameter.
Nonetheless, we decide to keep it and to check whether the perturbative corrections will actually turn out to be much smaller than $V_{LVS}$ in which case we claim our setup to be reliable.
Moreover, we notice that the largeness of $g_s$ is strictly connected to the fact that the D3-charge contributions coming from D-branes and O-planes is relatively small.
This is in turn due to the fact that, in order to have a simple construction, we have restricted our search to only those CY geometries for which $h^{1,1}<5$. 
However, we  expect to be able to improve this result once a larger number of \K moduli is taken into consideration.

The moduli VEVs for the uplifted LVS minimum are given as:
\bea
& & \langle \tau_5 \rangle =2.29 \, \, \, \, {\rm i.e.} \, \,  \langle t_5 \rangle =-0.76;\quad \langle \vol \rangle =2.05\times10^5,\,
\eea
which, substituted in the scalar potential \eqref{eq:V-LVS} give $V^{\rm LVS}_{\rm min}=5.12\times 10^{-18}$. We notice that all the constraints listed in \cite{Crino:2020qwk} and reviewed here in Section \ref{sec:list-of-requirements} are satisfied, with the caveat of the relatively large value of $g_s$ mentioned before.
In particular, the relevant energy scales of the model satisfy the required hierarchy:
\[M_s=3.3\times 10^{-3}m_p=6.5\,M_{KK}^{bulk}\:;\quad M_{KK}^{bulk}=5.1\times 10^{-4}m_p=51.3\, m_{3/2};\]
\[m_{3/2}=9.9\times 10^{-6}m_p\:.\]

\noindent
{Moreover we observe that the $2$-cycle volume modulus $t_5$ corresponding to the exceptional divisor is stabilized with a VEV such that $|\langle t_5 \rangle| < 1$. However, one may still hope that the corresponding corrections still remain under control and do not destabilize the minimum; for example, following the arguments from \cite{Denef:2005mm} where a similar situation appears after fixing the moduli, the largest worldsheet instanton corrections to the K\"ahler potential can be naively estimated to be,
\bea
& & \delta K \sim e^{-2\pi A} \sim 10^{-3},
\eea  
where we have used that the worldsheet-instantons wraps the exceptional curves with (string-frame) area $A = -2 \,\sqrt{g_s}\,\langle t_5 \rangle$.} As concerns the new constraints proposed in \cite{Junghans:2022exo}, our solutions satisfies all of them \textit{provided that} the complex structure depending constants are small enough (see the discussion at the end of Section \ref{sec:list-of-requirements}), while half of them are fulfilled, at least marginally, even for  $C_\circ^\circ=\mathcal{O}(1)$\footnote{We find, in particular: $\frac{\xi^{2/3}a_5^2 |\mathcal{C}_s^{log}|}{(2\kappa_5)^{2/3}g_s}=86 |\mathcal{C}_s^{log}|\,;\: \frac{2\xi^{1/3}a_5^2|\mathcal{C}_1^\xi|}{3(2\kappa_5)^{4/3}g_s}=162|\mathcal{C}_1^\xi|\,;\: \frac{2a_5 |\mathcal{C}_2^\xi|}{3(2\kappa_5)^{2/3}\xi^{1/3}}=12  |\mathcal{C}_2^\xi|\,;\: \lambda_1=0.14\,;\: \lambda_2=3.5\times 10^{-2}\mathcal{C}_b^{KK}\,;\: \lambda_3=6.8\,\mathcal{C}^{\rm flux}\,;\: \lambda_4=0.3\,\mathcal{C}^{\rm con}\,;\: \lambda_5=0.3\,\mathcal{C}^F\,,$ where we have highlighted the dependence on the complex-structure dependent constants. For the detailed origin and definition of these constraints, see \cite{Junghans:2022exo}.}. 

\noindent

\subsubsection{Stabilisation of the remaining moduli by sub-leading effects}
Since the $2$-cycle volume modulus $t_4$ appears linearly in the volume form \eqref{eq:ID1271vol}, it is convenient to write it in terms of the other $2$-cycle volumes (and the stabilised volume moduli in LVS): 
\beq
	t_4=-\frac{-8 t_5^3 + 3 t_2^2 t_8 + 3 t_2 t_8^2 - 8 t_8^3 + 6 \vol}{12 t_2 t_8 - 12 t_8^2}\:.
\eeq
Substituting this expression, as well as the values of the parameters and the moduli obtained in the previous step, we can analyse the resulting effective scalar potential in an iterative manner:
\[V=V^{\rm LVS}_{\rm min}+V_{g_s}+V_{F^4}\:.\]
{Subsequently we have scanned for dS solutions satisfying the conditions of Section~\ref{sec:list-of-requirements} by using the set of parameters $\{C_i^{KK}\}$ in the range $[10^{-4},1]$. One such solution with max$\{|C_i^{KK}|\} \sim 0.1$ is given as} \footnote{{We notice that the set of values for the $\{C_i^{KK}\}$ parameters resulting in a given dS solution, which satisfies the required constraints, are not always ${\cal O}(0.1-1)$, and at least one of those turns out to be ${\cal O}(10^{-3})$. However, such a sequestering in these parameters maybe attributed to the presence of highly warped throats; e.g. along the lines of \cite{Baumann:2010sx}.}}: 
\[C^{KK}_2=5.6\times 10^{-2};\quad C^{KK}_4=4.6\times 10^{-3}; \quad C^{KK}_5=10^{-3};\]
\[C^{KK}_8=-3.6\times 10^{-1}; \quad \lambda=-1.6\times 10^{-3},\]
for which the remaining moduli are fixed to: 
\[\langle t_2 \rangle =93.2; \qquad \langle t_8 \rangle =27.5,\]
and hence $ \langle t_4 \rangle = 91.9$.
The values of the parameters $C_i^{KK}$ and $\lambda$ have been chosen such that to ensure that the additional corrections to the scalar potential are actually sub-leading with respect to the LVS scalar potential.
We find, indeed, $\frac{V_{g_s}^{\rm KK}+V_{F^4}}{V_{\rm LVS + up}}=2.9\times10^{-2}$. 

As an aside, using Eq.~\eqref{eq:ID1271tau} we notice also that the stabilised value of the $4$-cycle $\tau_8$ turns out to be $\langle \tau_8 \rangle =3.1\times 10^{3}$; a non-perturbative correction coming from gaugino condensation on the divisor $D_8$ would be therefore suppressed by a factor $e^{-\frac{\pi}{3}\tau_8}/e^{-2\pi\tau_5}\simeq 10^{-1421}$ with respect to the one coming from the E3-instanton, therefore confirming the validity of our choice to neglect such an effect. Finally, we can compute the masses of the four \K moduli, considering the full scalar potential \eqref{eq:Vmain} along with the inverse K\"ahler metric which turns out to be given as below:
\[m_1^2=3.75\times10^{-8}m_p^2; \quad m_2^2=3.75\times 10^{-17}m_p^2;\]
\[m_3^2=3.29\times 10^{-19}m_p^2; \quad m_4^2=3.27\times 10^{-19}m_p^2.\]
Two of them are, as expected, much smaller than the others, while $m_1^2$ corresponds to the `small' cycle $\tau_5$.

\begin{figure}
	\includegraphics[width=.5\textwidth]{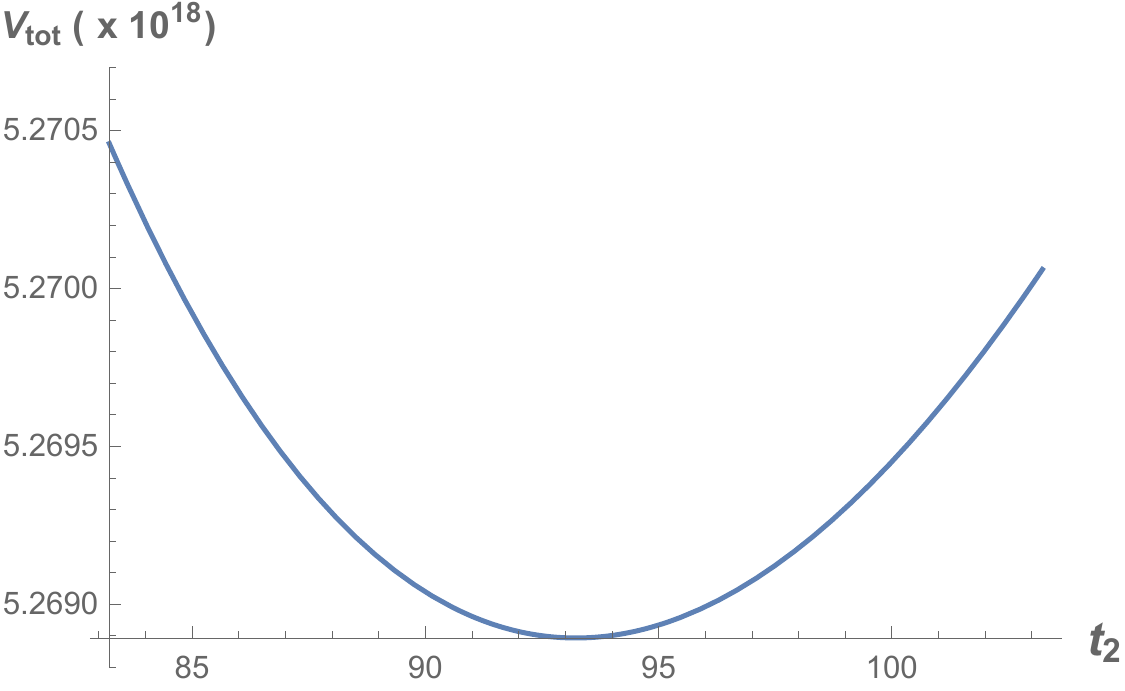}\includegraphics[width=.5\textwidth]{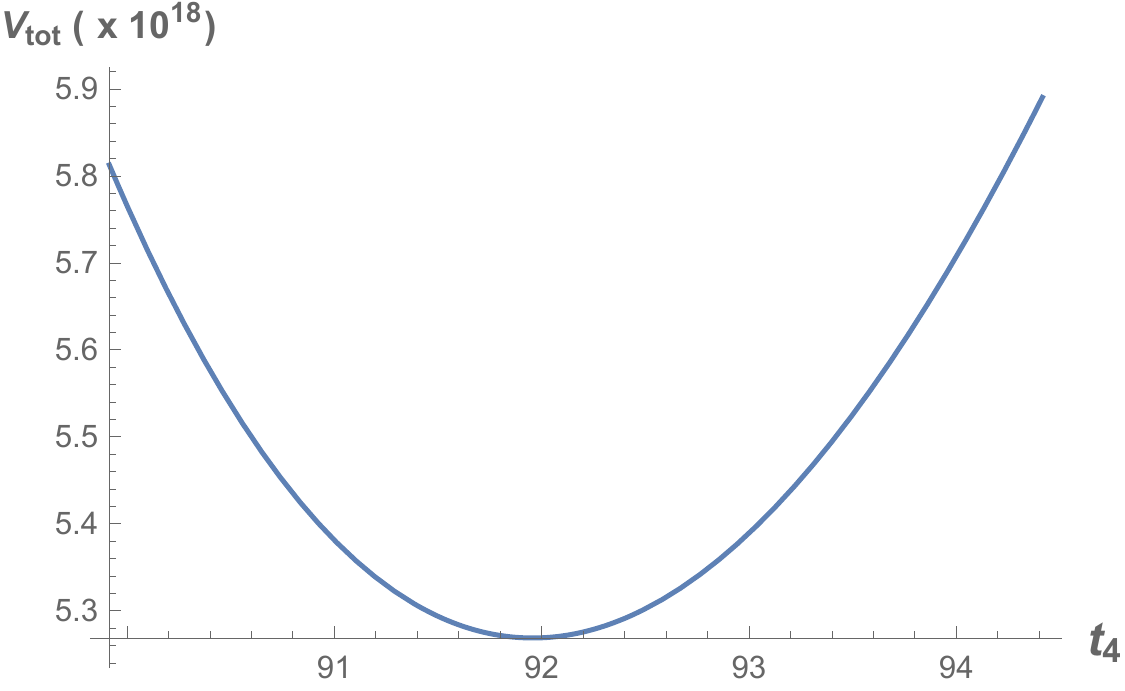}\\
	\includegraphics[width=.5\textwidth]{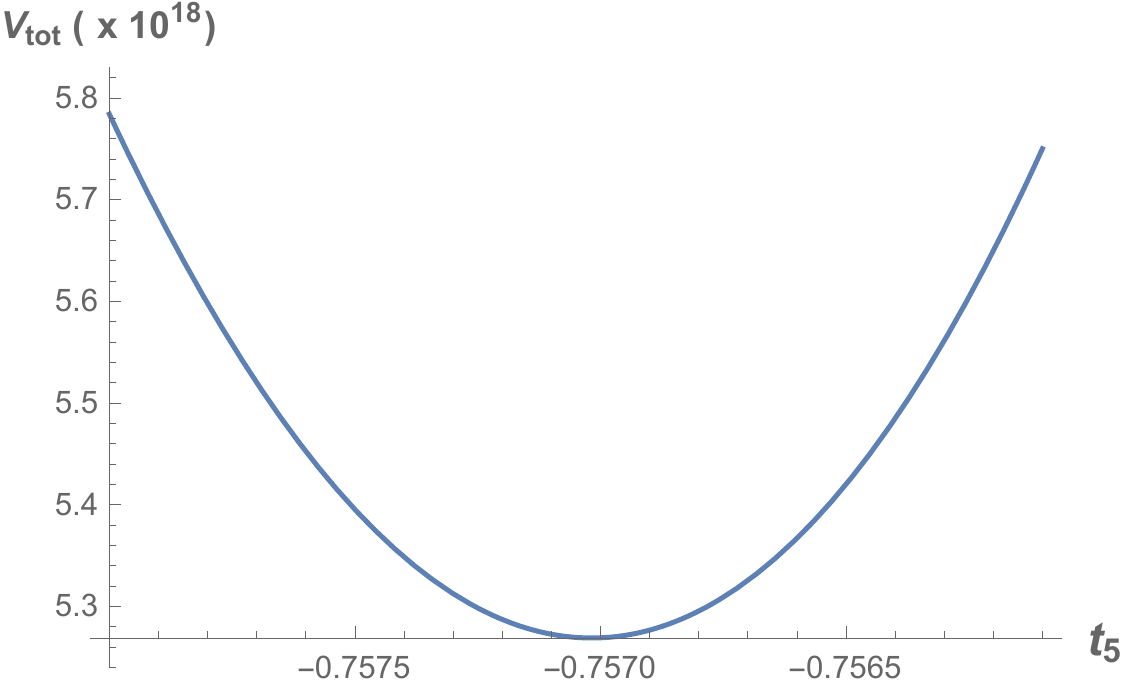}\includegraphics[width=.5\textwidth]{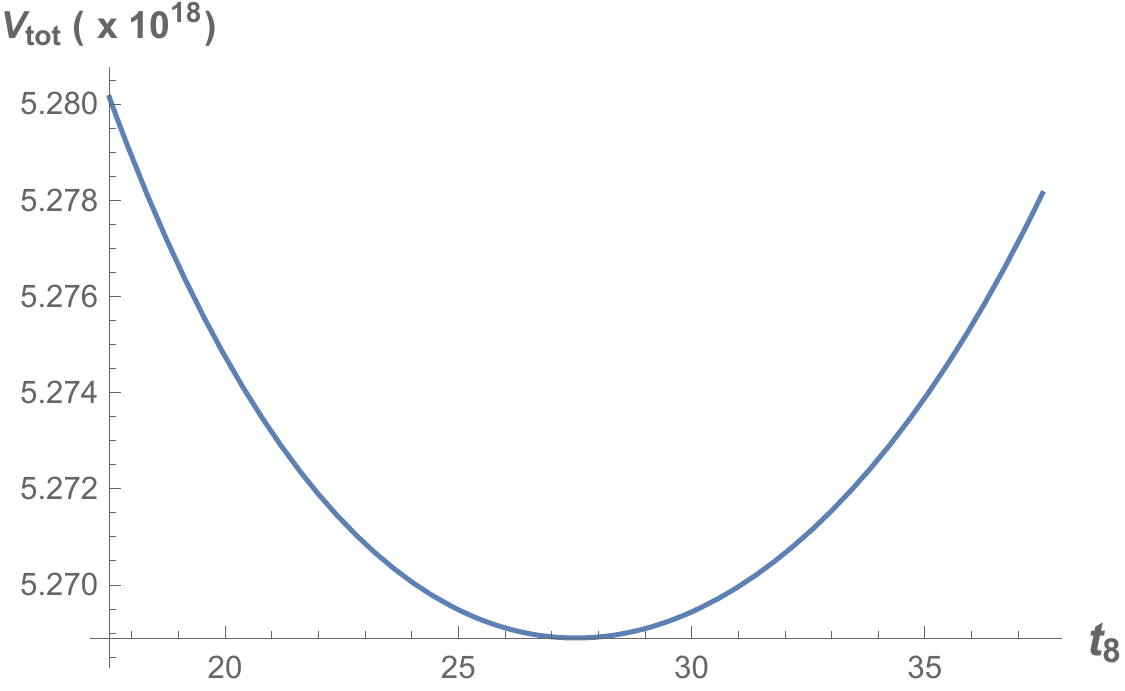}
	\caption{\label{fig:Vvst}Plots of the total scalar potential \eqref{eq:Vmain} along the four directions corresponding to the $4$ volume moduli $t^i$.}
\end{figure}

To summarise, our numerical model is characterized by the following parameters and moduli VEVs, corresponding to a dS minimum:
\bea
\label{eq:num-model}
& & |W_0|=4, \quad A_5 = 1, \quad g_s=0.52, \quad M=10 = K, \quad \chi_{\rm eff} =-166  , \quad C^{KK}_2=5.6\times 10^{-2} \nonumber\\
& & C^{KK}_4=4.6\times 10^{-3}, \quad C^{KK}_5=10^{-3}, \quad C^{KK}_8=-3.6\times 10^{-1}, \quad \lambda=-1.6\times 10^{-3}, \nonumber\\
& & \langle t_2 \rangle =93.2, \quad \langle t_4 \rangle = 91.9, \quad \langle t_5 \rangle =-0.76, \quad \langle t_8 \rangle =27.5, \quad \langle \vol \rangle =2.05\times10^5,\nonumber\\
& & \langle \tau_2 \rangle =2.1\times10^3, \quad \langle \tau_4 \rangle = 3.6\times10^3 , \quad \langle \tau_5 \rangle = 2.3, \quad \langle \tau_8 \rangle =3.1\times 10^{3}, \quad \langle V \rangle = 5.2\times 10^{-18} . \nonumber
\eea
The plots of the total scalar potential sliced along the directions corresponding to each of the $4$ volume moduli $t_i$, and in the vicinity of the VEVs corresponding to the actual minimum are shown in Fig. \ref{fig:Vvst}.


\section{Conclusion and future directions}
\label{sec:conclusion}

In this paper, we have added a further step towards the construction of phenomenologically interesting dS models in type IIB superstring compactifications, by taking into account a CY orientifold which, besides having all moduli stabilised in a dS minimum, possesses a K3-fibration structure. 

The combination of the new requirement with the ones to be imposed in order to have a setup which is suitable for the uplift mechanism based on \antiD branes, strongly constrains the admissible CY geometries at least for $h^{1,1}=3,4$.
After having explained the main reasons behind this obstruction, we have selected the best possibility in terms of the D3-charge and we have explicitly analysed it. 
In particular, after having presented its main geometric features, we have stabilised its four \K moduli in two separate steps, first by fixing à la LVS the overall volume of the CY and the volume of the $4$-cycle corresponding to the diagonal dP divisor. 
Then, in order to stabilise the remaining two moduli, we have introduced sub-leading corrections to the scalar potential arising from string-loop effects and the higher derivative $F^4$ corrections.

The resulting scalar potential allows for a good amount of analytic control, which we have used in order to guide the numerical analysis, performed by means of algorithms that could be useful in the future for the study of more involved scalar potentials, depending on several \K moduli. 
This is important, as we know that in order to get phenomenologically interesting constructions several \K moduli are needed and the corresponding scalar potentials become increasingly difficult to analyse. 
Our work, therefore, adds a further step to the several efforts that have been recently done in the attempt to analyse in a systematic manner models with an arbitrary number of \K moduli.
In the future, we would like to update this analysis in order to be able, in the end, to apply these algorithms to fully generic setups.

Our solution satisfies the constraints reported in Section \ref{sec:list-of-requirements}, with the caveat that $g_s$ is relatively large. 
The commonly considered bound for the string coupling is indeed $g_s<\frac{1}{3}$, while in our case we have $g_s\simeq\frac{1}{2}$. 
Nonetheless, we have checked that the additional corrections to the scalar potential used for stabilising the moduli which are left flat in the LVS step, are actually sub-leading for our setup, and therefore, on the basis of the hierarchical iterative nature of the various pieces, one would expect the overall mechanism to be trustworthy. 
Moreover, we anticipate that this value of the string coupling can be significantly improved by considering models with more than $4$ \K moduli, which should allow for larger (in absolute value) contributions to the D3-charge coming from O-planes and D-branes, hence for smaller values of $g_s$.

In our numerical analysis, we have taken into consideration also the recent constraints proposed in \cite{Junghans:2022exo}. However we think that requiring all of them to be fulfilled for $\mathcal{C}_\circ^\circ=\mathcal{O}(1)$ may be an overestimation, especially with our limited understanding/control of the explicit dynamics of all the moduli, including the complex structure and axio-dilaton in a single step. 
In particular, in our explicit setup we have allowed such constants, e.g. $\mathcal{C}_i^{\rm KK}$, to be smaller by a couple of orders, so that all the corrections considered in \cite{Junghans:2022exo} result to be numerically suppressed in our concrete global model.
Deriving some explicit mechanism to obtain such values is indeed beyond the scope of this paper, however recent proposals like \cite{Demirtas:2019sip} regarding realising exponentially low values for $W_0$ can indeed be considered as a caveat to the usual notion of what is really {\it tuned} and what is {\it natural},  with the concern becoming even milder when it is a matter of only a couple of orders of magnitude, and the complex structure moduli are implicitly present in the dynamics. However, it would be interesting to analyse these parameters in more detail in order to dynamically constrain their values using a generic scalar potential for all the moduli.

Finally, we stress that our choice of the values of the many parameters of the model was led by a systematic, but limited scan: more interesting choices might have escaped our analysis. 
Hence, a deeper analysis of the parameter space might be an interesting development for the future.


\acknowledgments

We thank Roberto Valandro for collaboration at the initial stage of this project. We would also like to thank Michele Cicoli, Andreas Schachner and Roberto Valandro for enlightening discussions. CC acknowledges support by INFN Iniziativa Specifica ST\&FI. PS would like to thank Paolo Creminelli, Atish Dabholkar and Fernando Quevedo for their support. This work was partly performed using resources provided by the Cambridge Service for Data Driven Discovery (CSD3) operated by the University of Cambridge Research Computing Service (www.csd3.cam.ac.uk), provided by Dell EMC and Intel using Tier-2 funding from the Engineering and Physical Sciences Research Council (capital grant EP/T022159/1), and DiRAC funding from the Science and Technology Facilities Council (www.dirac.ac.uk).


\bibliographystyle{JHEP}
\bibliography{bibliography}
%
\end{document}